\documentclass{article}
\usepackage{graphicx} 
\usepackage{subcaption}
\usepackage{amsmath}
\usepackage{amssymb}
\usepackage{fullpage}
\usepackage{xcolor}
\usepackage{color}
\usepackage{todonotes}
\usepackage[numbers]{natbib}
\usepackage[colorlinks=true,allcolors=blue]{hyperref}
\usepackage{csquotes}
\usepackage{booktabs}
\usepackage[normalem]{ulem}
\usepackage{array}

\title{Schema-based active inference supports rapid generalization of experience and frontal cortical coding of abstract structure}

\author{Toon Van de Maele$^{1}$ \and Tim Verbelen$^{1}$ \and Dileep George$^{2}$ \and Giovanni Pezzulo$^{3,*}$ \and  \\     
    $^1$ VERSES Research Lab, Los Angeles, USA \\
    $^2$ Google Paradigms of Intelligence, Mountain View, USA\\
    $^3$ Institute of Cognitive Sciences and Technologies, National Research Council, Rome, Italy\\
    $*$ Corresponding author: giovanni.pezzulo@istc.cnr.it
}

\date{\today}

\begin{document}

\MakeOuterQuote{"} 

\maketitle

\begin{abstract}
Schemas -- abstract relational structures that capture the commonalities across experiences -- are thought to underlie humans’ and animals’ ability to rapidly generalize knowledge, rebind new experiences to existing structures, and flexibly adapt behavior across contexts. Despite their central role in cognition, the computational principles and neural mechanisms supporting schema formation and use remain elusive. Here, we introduce schema-based hierarchical active inference (S-HAI), a novel computational framework that combines predictive processing and active inference with schema-based mechanisms. In S-HAI, a higher-level generative model encodes abstract task structure, while a lower-level model encodes spatial navigation, with the two levels linked by a grounding likelihood that maps abstract goals to physical locations. Through a series of simulations, we show that S-HAI generates candidate analogues to key behavioral signatures of rapid schema-based generalization in spatial navigation tasks, including the ability to flexibly remap abstract schemas onto novel contexts, resolve goal ambiguity, and balance reuse versus accommodation of novel mappings. Crucially, S-HAI also qualitatively recapitulates prominent neural codes reported in rodent medial prefrontal cortex during a schema-dependent navigation and decision task, including task-invariant goal-progress cells, goal-and-spatially conjunctive cells, and place-like codes at the lower level. Taken together, these results provide a candidate mechanistic account of schema-based learning and inference that bridges behavior, neural data, and theory. More broadly, our findings suggest that schema formation and generalization may arise from predictive processing principles, potentially implemented hierarchically across cortical and hippocampal circuits.

\end{abstract}

\textbf{Keywords:} schema; hierarchical active inference; predictive processing; frontal cortex, hippocampus

\section{Introduction}

Humans and other animals show a remarkable ability to rapidly generalize their knowledge to novel environments with very little new experience. This capability remains unparalleled by current artificial and AI systems, which usually require extensive datasets of problem-specific data.

In cognitive science, it has long been postulated that a specific cognitive structure—\emph{schemas}—may support the generalization of existing knowledge and skill to novel contexts. Schemas are typically defined as relational knowledge structures that capture abstracted commonalities across multiple experiences. They enable individuals to organize and interpret experiences in memory and to generalize to novel situations that share underlying structure, even when differing in sensory details \citep{piaget1952origins,bartlett1932remembering}. The crucial insight is that schemas are formed through experience, encoding inferred relational task structure while abstracting away from low-level (sensory) details. Schemas not only organize experiences into a rich set of relations but also serve as templates that enable the rapid assimilation of new experiences. This type of learning—or \emph{assimilation}—requires only mapping the low-level sensory details of a novel experience onto the abstract relational structure of an existing schema, enabling the fast (ideally, one-shot) reuse and generalization of knowledge to new contexts. As a result, it is significantly faster than the gradual accumulation of knowledge emphasized in classical theories of trial-and-error or associative learning. Furthermore, besides \emph{assimilation}—fitting new information into existing schemas—there is a second process—called \emph{accommodation}—which involves creating new schemas or modifying existing ones when new information does not fit. \citet{piaget1952origins} famously argued that this dual process of assimilation and accommodation is fundamental to learning and development.

These ideas originating in cognitive science have been influential since the early days of AI, inspiring numerous efforts to theorize and implement schemas (or related structures, such as frames or scripts), particularly but not exclusively within the symbolic AI tradition \citep{minsky1986society,schank2013scripts,hummel1997distributed}. Several theoretical accounts over the years have argued that the remapping of existing schemas and task representations to novel tasks lies at the heart of abstraction, structural inference, and analogy-making, as seen in multiple cognitive domains, from goal-directed navigation to rule learning and narrative understanding \citep{chollet2019measure,mitchell2021abstraction,hofstadter1999godel,tenenbaum2011grow,roy2005semiotic,pezzulo2009dipra,niv2019learning,schuck2016human,bein2025schemas,bahner2025abstract,collin2025neural,goudar2023schema,beukers2024blocked,yang2019task,sandbrink2024modelling}. However, designing effective computational models capable of learning schemas and generalizing them to novel experiences remains a significant challenge. Key difficulties include identifying the underlying relational structure to form schemas, encoding it in a form that supports rapid rebinding to new contexts, and designing mechanisms that enable rapid mapping between existing schemas and novel problems. Despite progress, a comprehensive computational account of schema-based learning and inference—particularly in complex, dynamic environments—remains elusive. Nonetheless, valuable insights into these challenges have begun to emerge from neuroscience.

In neuroscience, various studies have addressed the role of schemas in rodents and primates. These studies reveal that three interconnected brain structures—the hippocampus, the entorhinal cortex, and the frontal cortex—may play a crucial role in schema-based rapid learning and systems consolidation \citep{farzanfar2023cognitive}. For example, \citet{tse2007schemas} demonstrated that rats could integrate new information into an existing associative schema after only a single learning episode, with corresponding changes observed in hippocampal activity. This supports the view that schemas facilitate fast learning when new information is congruent with previously acquired relational structures. Additional studies have shown that hippocampal representations become more abstract and organized with schema acquisition, as evidenced by hippocampal replay and reactivation patterns during rest and sleep \citep{mckenzie2014hippocampal}. Beyond the hippocampus, grid cells in the entorhinal cortex have been implicated in cognitive map formation and schema learning \citep{neupane2024mental}. During spatial navigation, grid cells provide a periodic, low-dimensional representation of space that is believed to support path integration and map-like computations \citep{hafting2005microstructure}. More recent research suggests that grid-like coding may extend beyond physical space to support abstract cognitive maps, including task spaces and relational structures \citep{buzsaki2013memory,bellmund2018navigating,vigano2023mental,bottini2020knowledge,dong2024grid}. These studies suggest that grid cells in the entorhinal cortex may serve as a stable, reusable coordinate system that anchors task-specific details—mediated by the hippocampus—onto an abstract relational scaffold for organizing schematic knowledge.

Finally, various studies indicate that the frontal cortex—possibly in interaction with the hippocampus—plays a crucial role in forming cognitive maps and schematic associations, integrating new information into existing schemas that support generalization  \citep{gilboa2017neurobiology,zeithamova2012hippocampal,van2012schema,giuliano2021differential,bonasia2018prior,baldassano2018representation,basu2021orbitofrontal,manakov2025cognitive,schuck2016human,wang2021latent,vaidya2022abstract,samborska2022complementary,ElGaby2026}. For example, a study in which rodents navigated and generalized rules across distinct environments showed that this ability depends on coordinated hippocampal–prefrontal activity: the hippocampus formed environment-specific representations, whereas prefrontal population activity generalized across environments \citep{tang2023geometric}.

A particularly striking example of schema learning and generalization is provided by \citet{el2024cellular}, who demonstrated that the rodent medial frontal cortex is involved in learning the abstract structure of a sequential task—the ABCD task, which requires reaching four goal locations in the correct order—and in reusing this structure in novel environments where the sequence remains the same but the goal locations vary. A key indicator of fast, schema-based learning is that rodents rapidly move toward location A after discovering location D. At the neuronal level, this ability is supported by various types of cells that are sensitive to different combinations of abstract goal, physical location, and other task-relevant information. 

These and other studies have contributed to a convergent computational perspective, whose central insight is that during schema formation, relational structure is represented independently of sensory details. Each state in the relational structure is then bound to particular experiences via rapidly learnable associative links. This idea has been proposed in a series of computational models focusing on the hippocampus and the entorhinal cortex \citep{whittington2020tolman,whittington2018generalisation,whittington2025tale,chandra2025episodic}. In this view, the grid cell system in the entorhinal cortex provides the relational scaffold, while individual experiences are encoded in the hippocampus; the interaction between the entorhinal grid system and hippocampal place and conjunctive cells may allow rapid encoding of new experiences within a structured representational space, facilitating both assimilation and accommodation processes. Another related computational account of schema and rebinding in the hippocampus has been developed based on clone-structured causal graphs (CSCG) \citep{George2021,guntupalli2023graph,swaminathan2023schema, raju2024space}. In this perspective, the latent structure of the cognitive map of a maze can be abstracted by disconnecting it from the specific observations, and then reused as a schema to accelerate learning in other mazes. The same mechanism can also be used to learn the abstract structure of algorithms and template structures of language \citep{swaminathan2023schema} with inference time plasticity solves the problem of recalling the appropriate schema while dynamically binding the latent structure to novel inputs. Recent modeling work extends these ideas to neural activity in the frontal cortex but does not address how schemas are learned and deployed during spatial navigation \citep{el2024cellular}.

While valuable, these studies leave several important questions unanswered, including how abstract schemas can be formed and leveraged during goal-directed navigation and planning; how they extend to more challenging contexts in which multiple schemas—or multiple mappings between existing schemas and novel problems—must be created de novo, capturing the dynamics of what \citet{piaget1952origins} called assimilation and accommodation; and how they relate to neural processing in frontal cortex, as seen in the ABCD task \citep{el2024cellular}.

While prior work on CSCGs has addressed the problem of learning space as a latent structure from partially observable sensations \citep{raju2024space}, and CSCG schemas \citep{guntupalli2023graph, swaminathan2023schema} addresses the problem of transferring a learned spatial or algorithm structure using schemas, they do not address the problem of transferring multiple task structures that are learned in the same spatial environment. To do this in a partially observable setting, new task-structure schemas have to be learned that use the latent states of the spatial structure that was previously learned. 

In this study, we develop and validate a novel schema-based hierarchical active inference (S-HAI) model that aims to address these challenges. Here, a schema is represented as the dynamics (transition) model of a partially observable Markov decision process (POMDP), learned from experience and used for schema-guided inference and generalisation. For simplicity, our examples focus on the acquisition of a single schema and its reuse to generalize to novel problems (i.e., Piaget's \emph{assimilation}). This assimilation is enabled in our architecture by a specific component that rapidly remaps the abstract schema onto novel sets of stimuli, grounding it in the current situation (for this reason, we will later refer to this component as a \emph{grounding likelihood}). 

In the next section, we first introduce the experimental tasks (ABCD and ABCB, Section \ref{sec:tasks}) and the schema-based hierarchical active inference (S-HAI) agent that solves them (Section \ref{sec:S-HAI}). Then, we present four simulations that evaluate whether the S-HAI agent exhibits behavioral and neural (prefrontal) signatures of schema-based inference and learning reported empirically. The first simulation shows that schema-based inference in the S-HAI agent enables fast generalization in the ABCD task of \citet{el2024cellular} (Section \ref{sec:ABCD}). The second simulation demonstrates the efficacy of schema-based inference in a more demanding ABCB task, in which two goals can occupy the same location, as in spatial alternation tasks \citep{jadhav2012awake} (Section \ref{sec:ABCB}). The third simulation shows that the S-HAI agent is capable of incremental online learning and selection among multiple hypotheses about how to map its abstract schema onto the current maze (Section \ref{sec:mixture}). Finally, the fourth simulation shows that neural representations emerging in the S-HAI agent during learning of the ABCD task display key neural signatures of schema processing in the rodent mFC \citep{el2024cellular}.

\section{Results}

\subsection{Experimental tasks: the ABCD and ABCB tasks}
\label{sec:tasks}

\begin{figure}[t!]
    \begin{minipage}[b]{\textwidth}
        \begin{minipage}[b]{\textwidth}
            \centering
            \begin{minipage}[b]{0.45\textwidth}
                \centering
                \includegraphics[width=\textwidth]{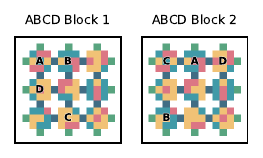}
                \subcaption[]{}
                \label{fig:taskABCD}
            \end{minipage}
            \begin{minipage}[b]{0.45\textwidth}
                \centering
                \includegraphics[width=\textwidth]{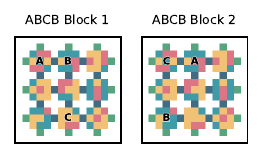}
                \subcaption[]{}
                \label{fig:taskABCB}
            \end{minipage}
        \end{minipage}
        \hfill
        \begin{minipage}[b]{.9\textwidth}
            \includegraphics[width=\textwidth]{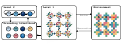}
            \subcaption[]{}
            \label{fig:hierarchicalmodel}
        \end{minipage}
        \centering
        \hfill
        \begin{minipage}[b]{0.19\textwidth}
            \centering
            \includegraphics[width=\textwidth]{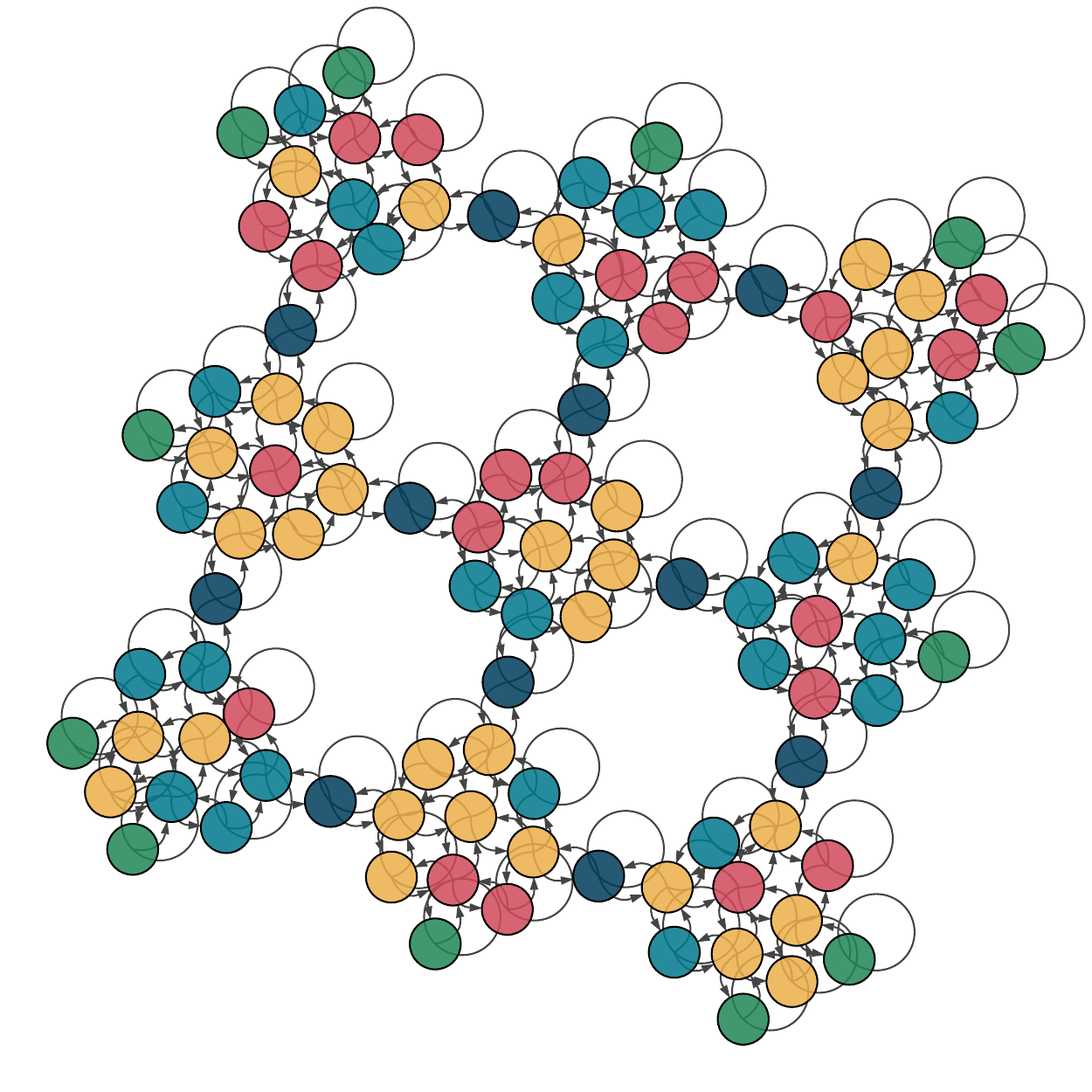}
            \subcaption[]{}
            \label{fig:navgraph}
        \end{minipage}
        \hfill
        \begin{minipage}[b]{0.19\textwidth}
            \centering
            \includegraphics[width=\textwidth]{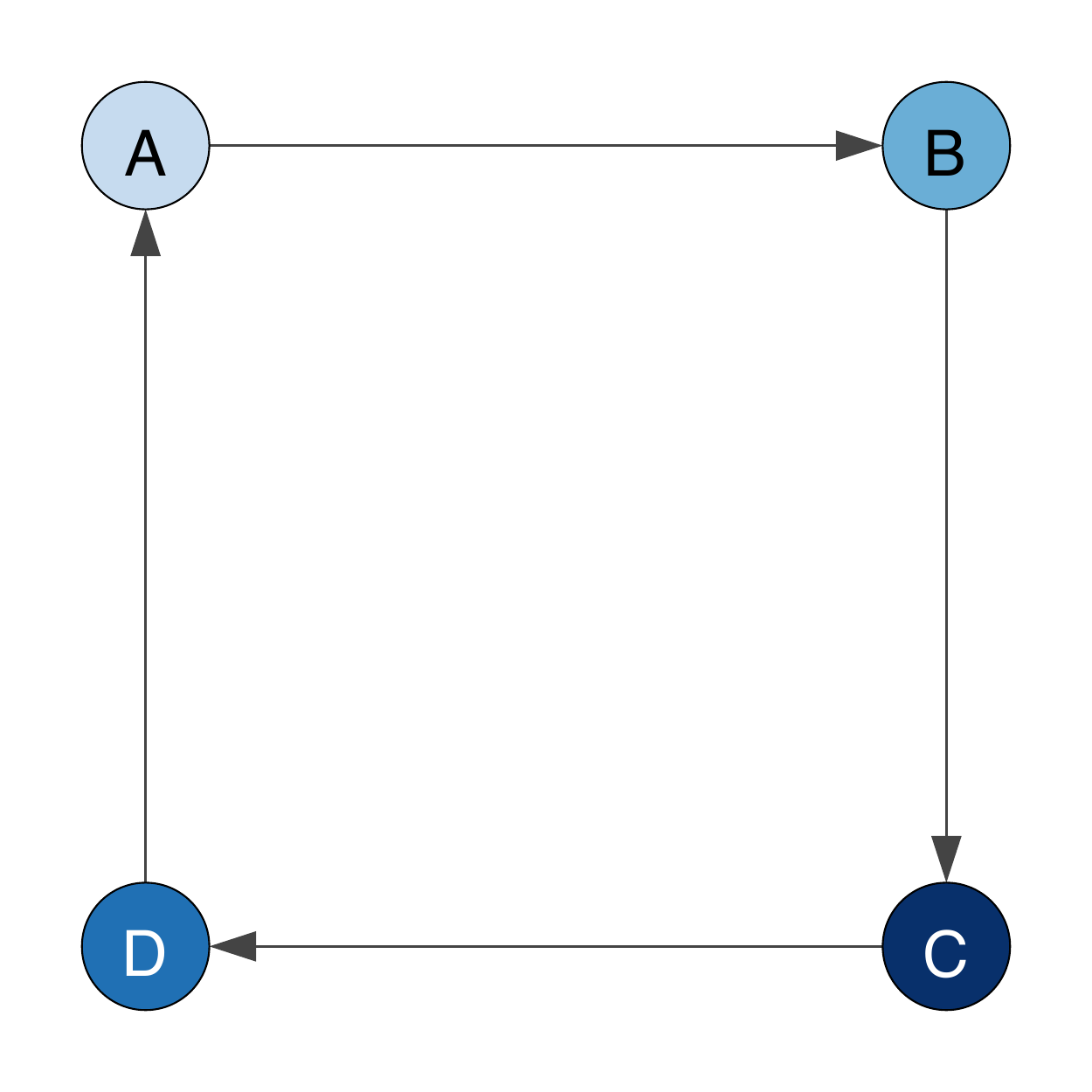}
            \subcaption[]{}
            \label{fig:task1}
        \end{minipage}
        \hfill
        \begin{minipage}[b]{0.19\textwidth}
            \centering
            \includegraphics[width=\textwidth]{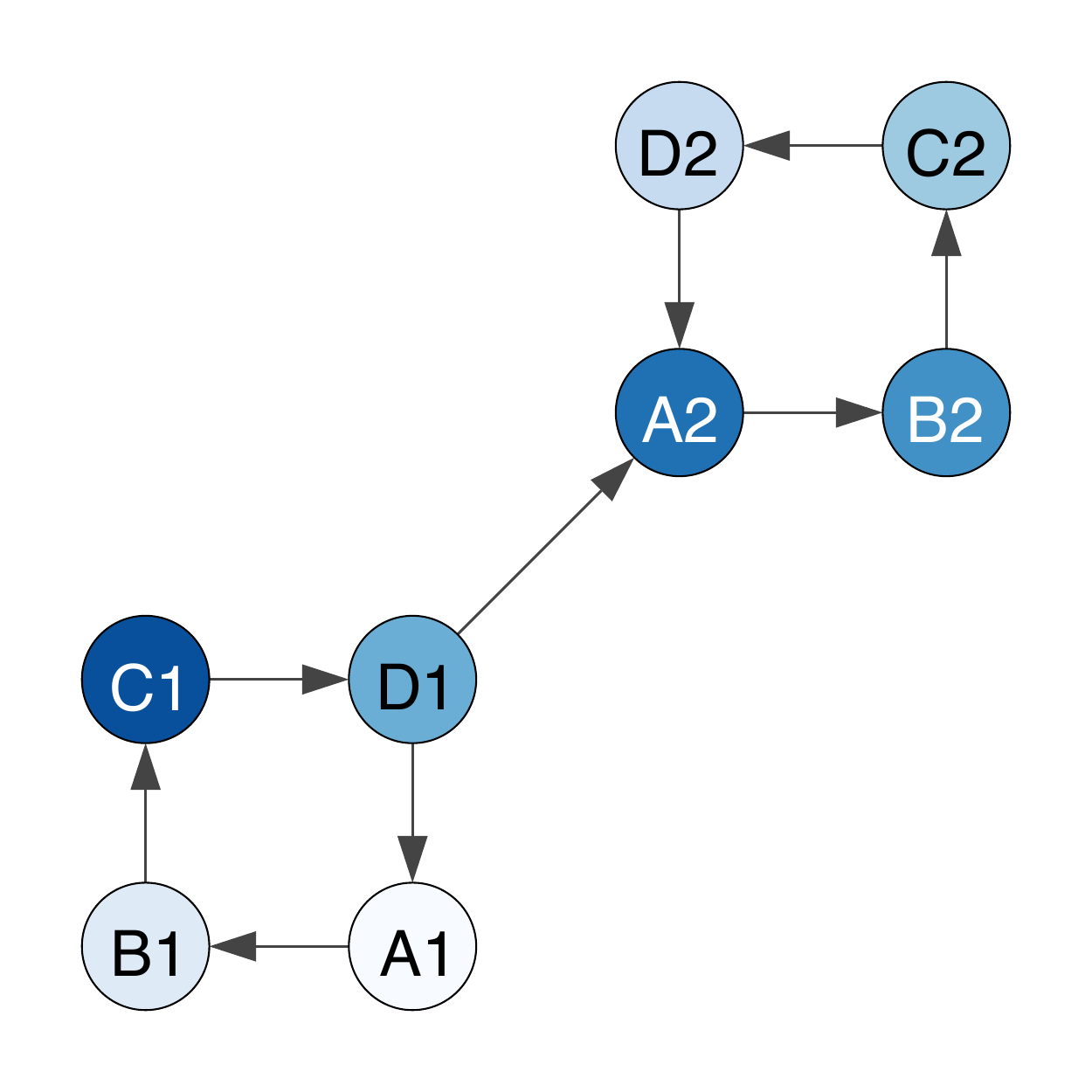}
            \subcaption[]{}
            \label{fig:task2}
        \end{minipage}
        \hfill
        \begin{minipage}[b]{0.19\textwidth}
            \centering
            \includegraphics[width=\textwidth]{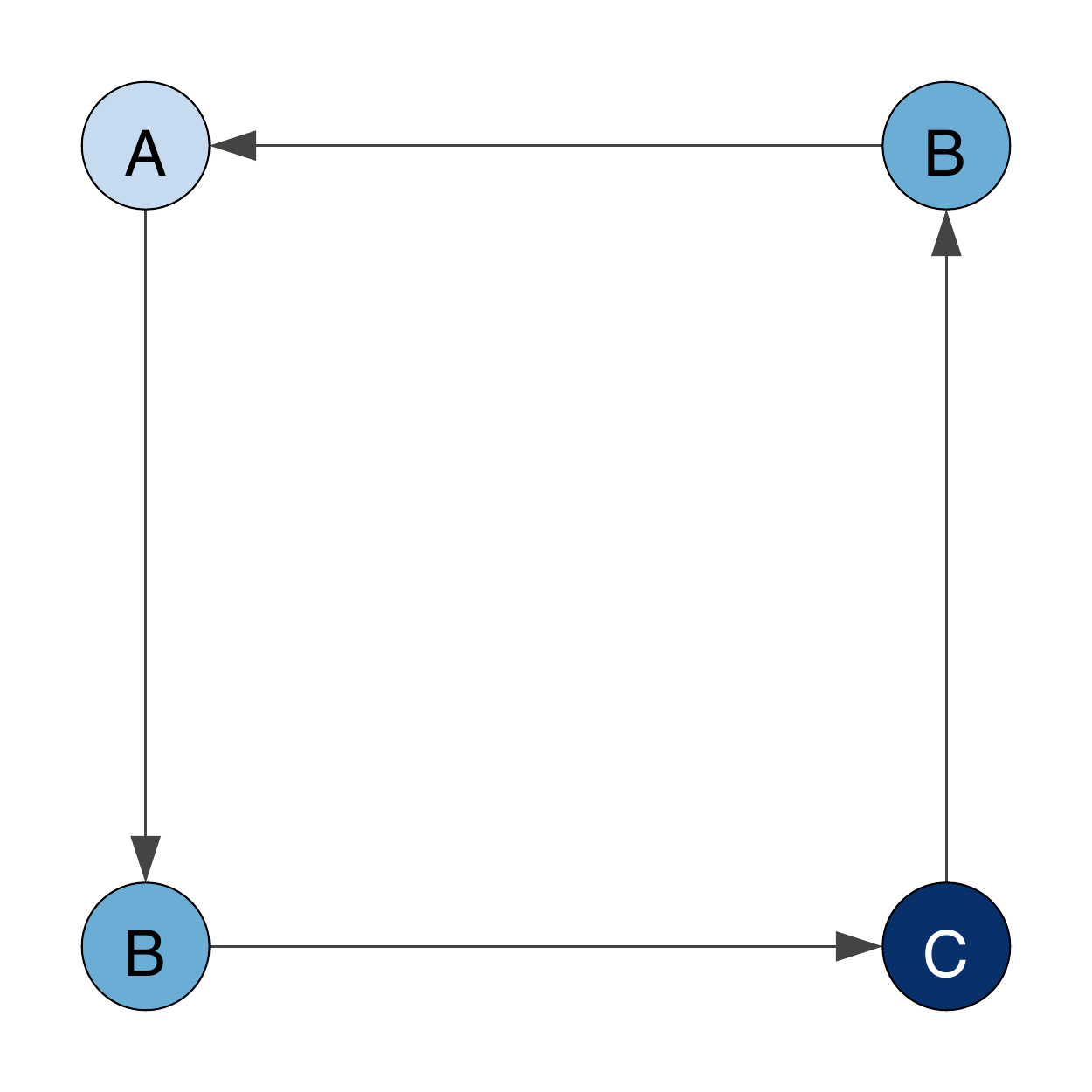}
            \subcaption[]{}
            \label{fig:task3}
        \end{minipage}

    \end{minipage}
    \caption{\textbf{Experimental tasks and schematic illustration of the schema-based hierarchical active inference model (S-HAI).} \small{ 
(a) Arrangement of the four goals in two example blocks of the ABCD task \citep{el2024cellular}. 
(b) Arrangement of the four goals in two example blocks of the ABCB task, requiring spatial alternation \citep{jadhav2012awake}. Note the two overlapping goals marked B.  
(c) Hierarchical generative model of the S-HAI agent. The figure depicts the interaction of the two levels of the model. High-level plans are formed in the task model (Level 2) and, through the grounding likelihood, are mapped onto the preferences of the spatial navigation model (Level 1). The agent then executes actions in the environment, which generate observations. Through bottom-up inference, beliefs over states are formed at each hierarchical level. The agent's location is marked in red. The tile color indicates the unique observations for the agent. The numbers indicate the specific Level 1 state they are bound to.    
(d) Transition matrix learned at Level 1 by the S-HAI model (trained on Block 1), encoding the spatial structure of the environment.  
(e) Transition matrix learned at Level 2 by the schema-based (S-HAI) agent trained on Block 1 of the ABCD task, representing transitions across the four goals A, B, C, and D.  
(f) Transition matrix learned at Level 2 by the agent without schemas (HAI) trained on Blocks 1 \& 2 of the ABCD task, representing transitions between the four goals of each block separately.  
(g) Transition matrix learned at Level 2 by the schema-based agent with two clone states (S-HAI-2C) trained on Blocks 1 \& 2 of the ABCB task.}}
\end{figure}

The main experimental task we adopt to evaluate our model is the ABCD task of \citet{el2024cellular}. In this task, a rodent (or artificial agent) acquires rewards by visiting four goal locations on a maze in the correct sequence (Figure \ref{fig:taskABCD}). The maze consists of nine wells arranged in a 3 $\times$ 3 grid, where each well consists of nine tiles which are connected through one-tile corridors. After the agent obtains a reward, a new reward is placed at the center of the next well in the sequence. The four goal locations differ across blocks, each comprising multiple trials. For example, in Block 1, the four goals are in maze locations: ``upper left'', ``upper center'', ``bottom center'', and ``center left'', whereas in Block 2, they are in locations: ``upper center'', ``bottom left'', ``upper left'', and ``upper right''. Crucially, the underlying (ABCD) structure remains invariant throughout the experiment: the rodent must always cycle through the four goal locations in the correct order (e.g., A, B, C, D, A, …). The distinction between variable sensory details and stable relational structure makes the ABCD task a natural testbed for schema-based inference.

We also address a more challenging variant, the ABCB task, in which two goals (the B goals) occupy the same spatial location (Figure \ref{fig:taskABCB}). This setup is similar to spatial alternation tasks commonly used in rodents \citep{jadhav2012awake} and is more demanding than the ABCD task, since at B animals must remember whether they arrived from A or from C in order to correctly select their next goal, C or A.

\subsection{Schema-based hierarchical active inference (S-HAI)}
\label{sec:S-HAI}

We address the ABCD and ABCB tasks using a novel schema-based hierarchical active inference (S-HAI) agent, which comprises two levels implemented as two interconnected Partially Observable Markov Decision Processes (POMDPs). Figure \ref{fig:hierarchicalmodel} provides a schematic illustration of S-HAI and its three components: Level 1, which handles spatial navigation; Level 2, which handles schema-based inference; and the \emph{grounding likelihood}, which specifies the probabilistic mapping between abstract goals encoded in schemas and particular locations in the maze. See Section \ref{sect:methods} for a formal specification of the S-HAI agent.


At the lower level (Level 1), the S-HAI agent handles spatial navigation in ``navigation space'', i.e., the grid world depicted in Figure~\ref{fig:taskABCD}. At this level, the agent observes the color of the tile it is currently visiting, and can navigate the grid using four actions (``up'', ``down'', ``left'', and ``right''). 
The parameters of the transition model used for navigation are obtained through offline training. In other words, we endowed the agent with the ability to navigate the map without knowing the task. This simulates the fact that, in the corresponding rodent studies, animals already know the environment but not the task before schema learning (see Section \ref{sect:methods}). Previewing the results of our simulations, we found that after training, the model correctly recovers the transition dynamics between the 105 locations (Figure \ref{fig:navgraph}, inset ``Level 1'' in Figure \ref{fig:hierarchicalmodel}).

At the higher level (Level 2), the agent performs schema-based learning and inference in ``task space''. Schema-based learning in the ABCD task amounts to learning an abstract transition model between goals, representing the fact that rewards are obtained by sequentially visiting four abstract goals, A, B, C, and D, then again A, and so on. Schema-based inference amounts to inferring the agent’s current position in task space (e.g., whether the current goal has been achieved) based on observations of Level 1 latent states and the presence or absence of reward, and then—if the current goal has been achieved—selecting the next navigation goal for Level 1. The Level 1 goal is specified as an intention over future state, which triggers the model to associate each state with an inductive cost~\citep{friston2023active}. This cost is proportional to the distance with respect to the intended goal state according to the latent dynamics (see Section~\ref{sect:methods} in Equation~\eqref{eq:inductivecost}). In our simulations, we implement both online and offline schema learning. Previewing our results, we found that in both cases, the S-HAI agent correctly learns a unique generalizable schema capturing the cyclical transitions between the four goals, applicable to both Blocks 1 and 2 (Figure \ref{fig:task1}, inset ``Level 2'' in Figure \ref{fig:hierarchicalmodel}). In contrast, an alternative agent trained without schema learning (HAI) captures distinct transitions for the two blocks (Figure \ref{fig:task2}). Finally, we found that an S-HAI agent augmented with the capability to distinguish between goals with identical locations (S-HAI-2C) correctly infers a generalizable schema for the ABCB task (Figure \ref{fig:task3}).

Crucially, the S-HAI agent also includes a \emph{grounding likelihood}: a probabilistic mapping between the abstract schema representing transitions between goals in task space (i.e., A, B, C, and D) and the concrete locations of the goals in navigation space (i.e., the locations in the maze where it can find reward). Learning grounding likelihoods is what affords schema-based generalization: it enables the agent to rapidly map its abstract schema (e.g., ABCD) to each novel spatial configuration of the goals (i.e., each block), rather than relearning the correct sequence of actions and goals from scratch on each trial. In our simulations, we implement online learning of the grounding likelihood, alongside both an online and offline variant for the Level 2 schema. In addition, we introduce a mixture model over grounding likelihoods that allows the S-HAI agent to flexibly infer which of its existing grounding likelihoods is most useful in the current maze, or to create a new one when needed. Previewing our results, we found that in all cases (online or offline, with or without mixture), grounding likelihoods allow the S-HAI agent to outperform alternative models trained on the same or even larger datasets but without schema learning. Learning the grounding likelihoods is similar to learning the emission matrix of a schema, as in \citep{guntupalli2023graph} and \citep{swaminathan2023schema}.

\subsection{The ABCD task: Schema-based hierarchical active inference allows fast generalization to novel problems with the same abstract structure}
\label{sec:ABCD}

In this simulation, we test whether schema learning enables the S-HAI agent to generalize the ABCD task across blocks of trials that share the same sequential structure but differ in specific goal locations. Following the experimental setting of \cite{el2024cellular}, each block ran until the agent completed 10K steps in the environment. A trial of four consecutive rewards can be completed in an average of $32 \pm 7.15$ steps $(\mu \pm \sigma)$, computed across 40 blocks.

We compare two variants of the S-HAI agent—one that learns schemas offline (S-HAI K) and one that learns them online (S-HAI L)—as well as a standard hierarchical active inference (HAI) agent without schemas and a baseline agent that selects goals randomly (Random). For consistency, in this and the subsequent simulations, all agents share the same hierarchical architecture. Furthermore, the Level 1 model responsible for spatial navigation (Figure~\ref{fig:navgraph}) is learned offline and is identical across agents. The only differences between agents arise at Level 2.  

In the offline schema-based agent (S-HAI K), the Level 2 schema is trained offline using data generated through a random walk (50 000 steps) collected from the first block only (the ``K'' denotes that the schema is known). Figure~\ref{fig:task1} visualizes the learned schema at Level 2, which represents a cycle between the four goals A, B, C, and D. The grounding likelihood is initialized randomly at the beginning of each block and learned online during the task. By contrast, in the online schema-based agent (S-HAI L), both the Level 2 schema and the grounding likelihood are trained online (the ``L'' denotes that the schema is learned online). The parameters are initialized randomly and updated using conjugate updates, with the grounding likelihood reset after each block.  

In the hierarchical active inference (HAI) agent without schemas, Level 2 is implemented using a clone-structured causal graph (CSCG) \citep{George2021}, and the grounding likelihood is an identity matrix. The agent is denoted as HAI-$i$, with the index $i$ indicating the number of tasks the agent is trained on; training is done offline on a sequence of the first $i$ tasks, with 10 000 interaction steps per task. See Figure~\ref{fig:task2} for the transition dynamics between goals learned by the HAI-$2$ agent (trained on two blocks) at Level 2. Note that, unlike the S-HAI agent, which learns a unique sequential schema, the HAI-$2$ agent learns a unique subcycle for each block. When the HAI-$i$ agent is trained on more blocks, it tends to learn multiple block-specific subcycles (not shown here, but similar to what is reported in~\citep{VandeMaele2024}).

Finally, in the baseline (Random) hierarchical active inference model, Level 2 randomly selects a subgoal for the agent to navigate to. See Section \ref{sect:methods} for a formal explanation of the agents used in this simulation.

\begin{figure}[t!]
    \centering
    \begin{minipage}{0.48\linewidth}
        \centering
        \begin{minipage}[b]{\textwidth}
           \centering 
           \includegraphics[width=0.9\linewidth]{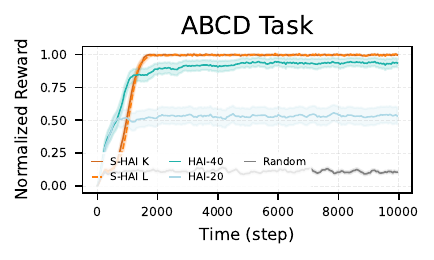}
           \subcaption[]{}
           \label{fig:abcd_reward}
        \end{minipage}
        \begin{minipage}[b]{0.47\textwidth}
           \centering 
           \includegraphics[width=\linewidth]{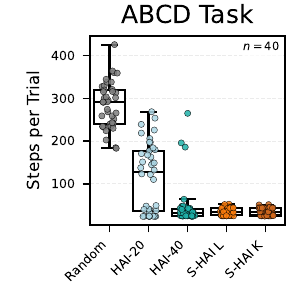}
           \subcaption[]{}
           \label{fig:timesboxplot_abcd}
        \end{minipage}
        \hfill
        \begin{minipage}[b]{0.47\textwidth}
           \centering 
           \includegraphics[width=\linewidth]{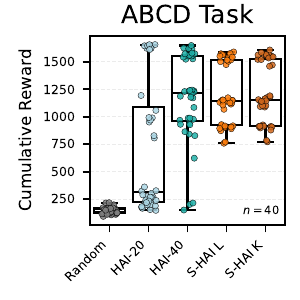}
           \subcaption[]{}
           \label{fig:rewardsboxplot_abcd}
        \end{minipage}
        \begin{minipage}[b]{0.47\textwidth}
           \centering 
           \includegraphics[width=\linewidth]{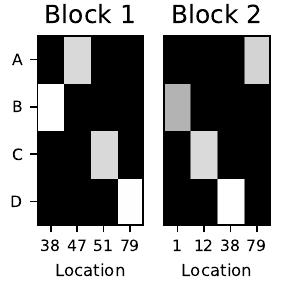}
           \subcaption[]{}
           \label{fig:remap_abcd}
        \end{minipage}
        \hfill
        \begin{minipage}[b]{0.47\textwidth}
            \vspace{1em}
           \centering 
           \includegraphics[width=\linewidth]{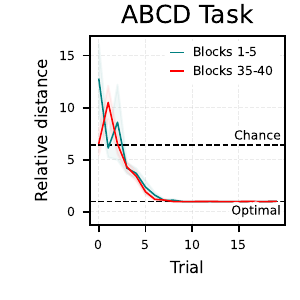}
           \subcaption[]{}
           \label{fig:firstlast_abcd}
        \end{minipage}
    \end{minipage}
    \begin{minipage}{0.47\linewidth}
        \centering
        \begin{minipage}[b]{\textwidth}
           \centering 
           \includegraphics[width=0.9\linewidth]{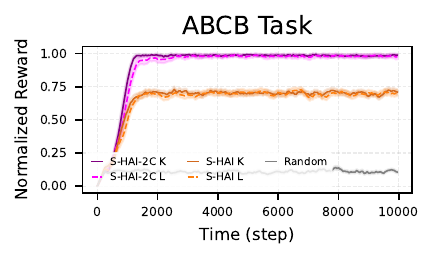}
           \subcaption[]{\label{fig:abcb_reward}}
        \end{minipage}
        \begin{minipage}[b]{0.47\textwidth}
           \centering 
           \includegraphics[width=\linewidth]{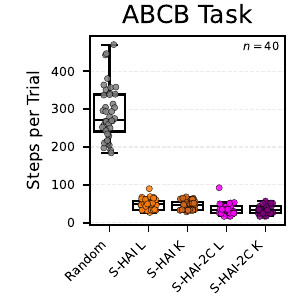}
           \subcaption[]{\label{fig:timesboxplot_abcb}}
        \end{minipage}
        \begin{minipage}[b]{0.47\textwidth}
           \centering 
           \includegraphics[width=\linewidth]{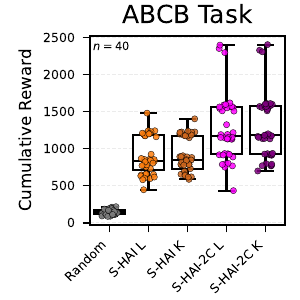}
           \subcaption[]{\label{fig:rewardsboxplot_abcb}}
        \end{minipage}
        \begin{minipage}[b]{0.47\textwidth}
           \centering 
           \includegraphics[width=\linewidth]{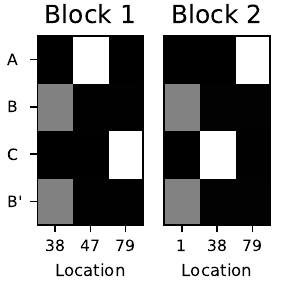}
           \subcaption[]{\label{fig:remap_abcb}}
        \end{minipage}
        \begin{minipage}[b]{0.47\textwidth}
            \vspace{1em}
           \centering 
           \includegraphics[width=\linewidth]{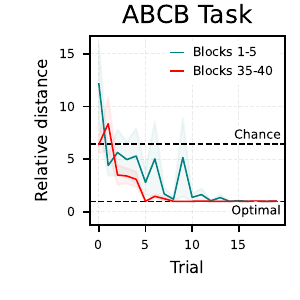}
           \subcaption[]{\label{fig:relative_abcb}}
        \end{minipage}
    \end{minipage}
    \caption{\textbf{Simulations 1 and 2: addressing the ABCD and ABCB tasks, with and without schemas.} 
\small{(a) Normalized reward in the ABCD task for the two schema-based agents using offline (S-HAI K) and online (S-HAI L) learning; two non-schema agents trained offline on 20 (HAI-20) or 40 blocks (HAI-40); and baseline (Random) agent. Solid line shows mean normalized reward (measured as the reward rate smoothed over 250 steps, normalized against optimal performance) across 40 blocks; the shaded area shows the standard error. Each block ends when the agent reaches a maximum of 10 000 navigation steps.
(b) Box plot of the average number of steps per trial across 40 blocks required to complete an ABCD task and obtain 4 consecutive rewards, for the various agents.  
(c) Box plot of cumulative rewards across 40 ABCD blocks, for the various agents.  
(d) \emph{Grounding likelihoods} for the first two blocks of the ABCD task learned by the S-HAI L agent. Values range between 0 (black) and 1 (white).  
(e) Relative distance to subgoals (ratio of path length taken to shortest possible path between subgoals) in first and last 5 blocks of the ABCD task, for the S-HAI L agent.  
(f) Performance in the ABCB task for offline and online schema-based agents with clone graphs (S-HAI-2C K, S-HAI-2C L), without clone graphs schema-based agents (S-HAI K, S-HAI L), and the Random agent.  
(g) Box plot of average number of steps across 40 blocks per trial required to complete ABCB task and obtain 4 consecutive rewards, for the various models.  
(h) Box plot of cumulative rewards across 40 ABCB blocks, for the various models.  
(i) \emph{Grounding likelihoods} for the first two blocks of the ABCB task learned by the S-HAI-2C K agent.  Values range between 0 (black) and 1 (white).      
(j) Relative distance to subgoals in the first and last 5 blocks of the ABCB task, for the S-HAI-2C L agent. Statistical significances in Tables \ref{tab:timesboxplot_abcd_effectsize}, \ref{tab:rewardsboxplot_abcd_effectsize}, \ref{tab:timesboxplot_abcb_effectsize}, and \ref{tab:rewardsboxplot_abcb_effectsize}.}}
\label{fig:behavioral}
\end{figure}

Figure~\ref{fig:abcd_reward} shows the simulation results for the ABCD task, reporting the average reward rate (smoothed over 250 steps), normalized against optimal performance, across 40 blocks. The solid line denotes the mean across blocks, and shaded areas indicate the standard error. Each block ends when the agent reached 10 000 interactions with the environment. As expected, the HAI agents without schemas (trained offline on half (20) or all (40) blocks; HAI-20, and HAI-40) outperformed the Random baseline, with performance improving as the number of training blocks increased. However, the HAI-20 agent trained in half of the blocks showed limited generalization to novel blocks beyond its training set (see Appendix~\ref{app:seenunseen}, Figure~\ref{fig:seenunseen}).  

In contrast, the schema-based S-HAI agents generalized robustly to unseen tasks. The offline agent (S-HAI K) rapidly converged to near-optimal reward levels, demonstrating that a schema learned from a single training block can generalize to 39 novel blocks. The online agent (S-HAI L) also reached near-optimal performance, showing that generalizable schemas can be learned efficiently online without prior offline training (see Appendix~\ref{app:a2t} for an analysis quantifying the acquisition-to-transfer advantage conferred by using a learned schema upon first exposure to novel blocks. Both S-HAI agents required fewer steps per trial (Figure~\ref{fig:timesboxplot_abcd}) and accumulated more reward (Figure~\ref{fig:rewardsboxplot_abcd}) than the HAI-20 agent and the Random baseline (see Supplementary Tables \ref{tab:timesboxplot_abcd_effectsize} and \ref{tab:rewardsboxplot_abcd_effectsize}). Additionally, both S-HAI agents achieve the same level of average performance as the extensively trained HAI-40 agent (Figures~\ref{fig:abcd_reward},\ref{fig:rewardsboxplot_abcd},\ref{fig:timesboxplot_abcd}). A more detailed analysis shows that both S-HAI agents are more flexible in adapting to variations of the ABCD task, successfully learning all 40 blocks, whereas HAI-40 fails to converge in 3/40 blocks, plausibly because the goal configurations in these blocks can be confused with those of other blocks (see Appendix~\ref{app:seenunseen}, Figure~\ref{fig:seenunseen}). 
To ensure that the observed behavior is not reflective of imperfect learning at the lower level, we also replicated our results in a simplified environment (a $3 \times 3$ grid, with unique observations; see Appendix~\ref{app:smallmaze}). 

A key driver of schema-based generalization is that both S-HAI K and S-HAI L learned a new \emph{grounding likelihood} online in each block, mapping abstract schema observations ($\mathbf{o}^2_t$) to concrete spatial states ($\mathbf{s}^1_t$). Two examples of grounding likelihoods are shown in Figure~\ref{fig:remap_abcd}, where only the high-level states corresponding to rewarding locations are displayed for clarity (the full grounding likelihood includes 210 high-level states: the conjunction of 105 locations with reward presence). Schemas and grounding likelihoods were acquired quickly within blocks, as indicated by the rapid improvement of S-HAI L performance with experience, reflected in the reduction of relative distance to subgoals (Figure~\ref{fig:firstlast_abcd}).  

Summing up, we found that schema-based S-HAI agents successfully solve structured tasks such as ABCD by rapidly grounding abstract, schema-encoded goals in physical locations that vary across blocks. Remarkably, these schema-based agents perform far more efficiently than the HAI-20 agent trained on only half of the blocks, and reach maximal reward faster than the HAI-40 agent, despite the latter being extensively trained offline on all block configurations. This result suggests that, in structured tasks such as ABCD, schema-based generalization can be as efficient as exhaustive offline learning while being even more flexible to variations on the task.


\subsection{The ABCB task: Schema-based hierarchical active inference augmented with clone graphs can address problems with aliased goals}
\label{sec:ABCB}

In this simulation, we consider a more challenging variant of the ABCD task that includes an alternation pattern between goals \citep{jadhav2012awake}. Here, the second and the fourth goals, both denoted as B, occupy the same spatial location; this is why the task is called ABCB.  

What makes this task more difficult is that it requires spatial memory: when the agent observes a reward at location B, it must decide whether to move toward the C or the A target. Standard HMM-like architectures, such as the one used by the HAI agent in the first simulation, struggle with this task because they confuse the two instances of the B goal. To address this limitation, we endowed Level 2 of the HAI agent with a more expressive clone-structured causal graph (CSCG) mechanism \citep{George2021}, which extends HMMs by allowing multiple clones for each state. We call the resulting agent S-HAI-2C K. Here, “2C K” denotes that Level 2 is a CSCG with two clones, trained offline using a random walk (10 000 steps) from the first block, which we found sufficient to learn the ABCB schema (Figure~\ref{fig:task3}). Additionally, we include a variant that learns the parameters of the clone-structured causal graph at Level 2 online, called S-HAI-2C L. As in the first simulation, the grounding likelihood was initialized randomly at the beginning of each task and learned online during the task. See Section~\ref{sect:methods} for a formal specification of the S-HAI-2C K agent.  

Figure~\ref{fig:abcb_reward} shows the mean acquired reward over time across 40 blocks of the ABCB task. The results show that the schema-based agents with clones (S-HAI-2C K, S-HAI-2C L) reached near-optimal performance, outperforming the two agents without clones (S-HAI K, S-HAI L), which struggled with the ambiguity of the B goal. As shown in Figures~\ref{fig:timesboxplot_abcb} and \ref{fig:rewardsboxplot_abcb}, the S-HAI-2C K agent requires fewer steps per trial and achieves higher reward than its non-clone counterparts.

It is also interesting to note that, for the ABCB task, in the first blocks, the relative distance (Figure~\ref{fig:relative_abcb}) for each trial decreases more slowly than during the late trials, indicating that in the late stage, the agent has learned the abstract task structure in the schema and only has to infer the grounding likelihood. 


Figure~\ref{fig:remap_abcb} shows the grounding likelihoods for two blocks learned by the S-HAI-2C K agent. At Level 2, the grounding likelihood $P(\mathbf{s}^1_t | \mathbf{o}^2_t)$ is combined with the observation likelihood $P(\mathbf{o}^2_t|\mathbf{s}^1_t)$. Unlike Figure~\ref{fig:remap_abcd}, two distinct task states (corresponding to the first and second occurrences of goal B) map to the same spatial locations (location 2 in Block 1 and location 7 in Block 2). 

Taken together, these findings illustrate that augmenting the S-HAI model with clone-based mechanisms \citep{George2021} enables it to learn schemas that generalize effectively in tasks with aliased goals. 

\subsection{Schema-based inference with a mixture model affords the incremental learning and reuse of multiple grounding likelihoods}
\label{sec:mixture}

In the previous simulations, we examined how the agent remaps spatial states to task states using a single grounding likelihood, which was re-trained at each block. However, in real-world scenarios, animals (and artificial agents) may need to autonomously identify when blocks of problems change \citep{behrens2007learning,friston2016active,proietti2025active}. Furthermore, they may encounter the same blocks multiple times, in which case re-learning them from scratch would be inefficient.  

To address this challenge, we implemented a non-parametric extension of the S-HAI agent, called the S-HAI MoGL agent, which maintains a mixture of grounding likelihoods that expand over time using a truncated Dirichlet process (Section~\ref{sect:methods}). The S-HAI MoGL agent maintains a belief over the mixture, which is reset to a uniform prior at the beginning of each block, and selects the most likely grounding likelihood for each trial. This non-parametric approach allows the agent to flexibly create new grounding likelihoods when encountering novel problems, while reusing existing grounding likelihoods when encountering previously seen problems. As in the previous simulations, the non-parametric agent was implemented with both online learning (S-HAI L MoGL) and offline learning (S-HAI K MoGL). See Section~\ref{sect:methods} for a formal explanation of the S-HAI MoGL agent.  

\begin{figure}[tb!]
    \centering
    \begin{minipage}{0.90\linewidth}
        \centering
        \begin{minipage}{\linewidth} 
            \centering
            \includegraphics[width=0.85\linewidth]{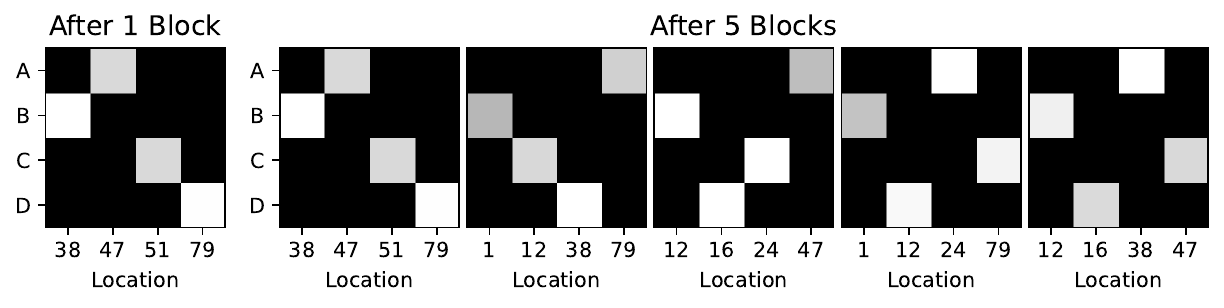}
            \subcaption[]{\label{fig:remappings}} 
        \end{minipage}
        \begin{minipage}{0.45\textwidth}
            \includegraphics[width=0.95\linewidth]{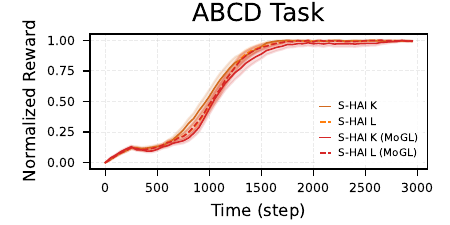} 
            \subcaption[]{\label{fig:mix_abcd}}
        \end{minipage}
        \begin{minipage}{0.45\textwidth}
            \includegraphics[width=0.95\linewidth]{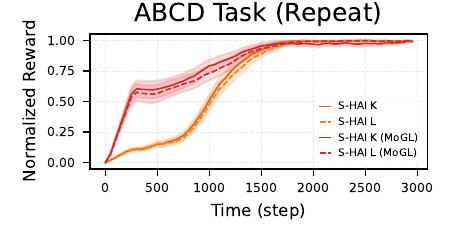} 
            \subcaption[]{\label{fig:mix_abcd_repeat}}
        \end{minipage}
        \begin{minipage}{\linewidth}
            \centering
           \begin{minipage}{0.45\linewidth}
                \includegraphics[width=0.95\linewidth]{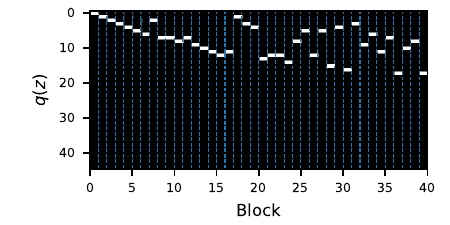} 
                \subcaption[]{}
                \label{fig:moglqz}
           \end{minipage} 
           \begin{minipage}{0.45\linewidth}
                \includegraphics[width=0.95\linewidth]{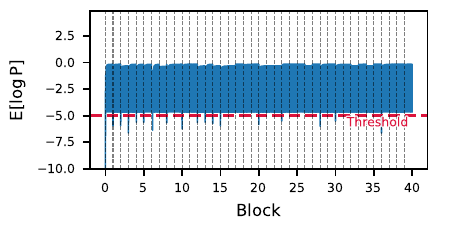} 
                \subcaption[]{}
                \label{fig:moglell}
           \end{minipage} 
        \end{minipage}
    \end{minipage}
    \caption{\textbf{Simulation 3: addressing the ABCD task with schemas and a mixture of grounding likelihoods.} \small{
(a) Grounding likelihoods learned by the S-HAI MoGL L agent after the first block (left) and after five blocks (right) of the ABCD task. For visual clarity, the grounding likelihoods only show the relevant entries. The location indicates the specific value for the Level 1 states across the blocks. Values range between 0 (black) and 1 (white).
(b) Normalized reward over 3000 timesteps on the ABCD task, with 40 sequentially experienced blocks. The shaded area shows the standard error.
(c) Normalized reward over 3000 timesteps on a variant of the ABCD task with repeated blocks. In this variant, the first 20 blocks are randomly sampled to generate 40 blocks, so that some blocks can be repeated multiple times. The shaded area shows the standard error.  
(d) The approximate posterior over $z$ (the selector variable of the grounding likelihood) during the ABCD task with repeated blocks, showing that the agent either creates a new grounding likelihood or reuses an existing one when it re-encounters an environment. 
(e) The expected log likelihood during the 40 blocks of the ABCD task with repeated blocks, with the threshold (-5) used for expanding the mixture model. See the main text for details. }}
    \label{fig:enter-label}
\end{figure}

To test the mixture of grounding likelihoods, we tested the S-HAI L MoGL and S-HAI K MoGL agents in the ABCD task used in the first simulation, where the agent faces 40 distinct blocks of problems. Our results show that after completing the first block, the S-HAI MoGL agents learn a single grounding likelihood, shown on the left of Figure~\ref{fig:remappings}. As the agent encounters novel blocks, the mixture model expands, resulting in multiple distinct grounding likelihoods for different blocks; for example, the right of Figure~\ref{fig:remappings} shows the learned mixture components after five blocks.

Figure~\ref{fig:mix_abcd} compares the performance of schema-based agents with (S-HAI MoGL K and S-HAI MoGL L) and without (S-HAI K and S-HAI L) a mixture of grounding likelihoods in the ABCD task. The results for the S-HAI K and S-HAI L agents without a mixture are identical to those shown in Figure~\ref{fig:abcd_reward}. Our results show that introducing a mixture of grounding likelihoods affects how quickly the S-HAI MoGL agent learns each task. The S-HAI MoGL agent—which must infer which grounding likelihood applies to the current problem and, in some cases, create a new one—has a similar reward curve to a schema-based agent that retrains a single likelihood at each new block (see Appendix~\ref{app:convergence}, Tables~\ref{tab:stepstocriterion_abcd_effectsize}~and~\ref{tab:stepstocriterion_abcdrepeat_effectsize}).  

Finally, to further examine the benefits of the mixture of grounding likelihoods, we consider a variant of the ABCD task in which the agent can encounter the same blocks multiple times. In this variant, the 40 blocks that compose the experiment are random samples of only the first 20 blocks used in the previous simulations, so certain blocks may be repeated twice or more. Figure~\ref{fig:mix_abcd_repeat} shows the results. Both the schema-based agents with a mixture of grounding likelihoods (S-HAI MoGL) and with a single grounding likelihood (S-HAI) eventually reach the same optimal performance. However, the agent endowed with the mixture model learns faster, as it can immediately apply previously acquired components to tasks it has already encountered. This simulation illustrates that maintaining multiple explicit mappings between the schema and the problems in which it can be applied provides a clear advantage in tasks where prior knowledge can be reused. 

Figure~\ref{fig:moglqz} provides a more detailed view of how the S-HAI MoGL agent accumulates and organizes the mixture of grounding likelihoods during the ABCD task with repeated environments. The panel shows the agent’s belief over the selected mixture component (i.e., the grounding likelihood) across time. In most cases, the agent assigns a unique grounding likelihood to each block, reflecting successful differentiation of the blocks. The bottom panel shows the expected log likelihood of the observations under the mixture model (the first factor in \eqref{eq:qz}). When this quantity falls below a threshold (shown as the dotted red line in Figure~\ref{fig:moglell}), a new grounding likelihood is added to the mixture. This typically occurs at the beginning of a new block, when observations are surprising and inconsistent with prior expectations—consistent with empirical findings that boundaries between episodes often correspond to moments of high surprise \citep{zacks2020event}. This does not happen when a block is encountered that the agent has previously observed. In the ABCD task without repeating environments, the agent consistently creates new mixture components for each encountered block (see Appendix~\ref{app:mogl}, Figure~\ref{fig:mogl_norepeat}). 

\subsection{Schema-based hierarchical active inference generates analogous patterns to ``goal-progress cells'' and other key signatures of schemas in the rodent medial frontal cortex}

\begin{figure}[t!]
    \centering
    \begin{minipage}[b]{0.49\linewidth}
        \includegraphics[width=\linewidth]{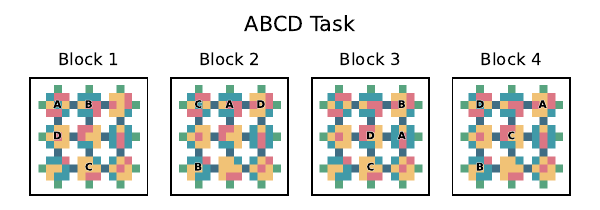}
        \subcaption[]{\label{fig:tasks_neural}}
    \end{minipage}
    \begin{minipage}[b]{0.49\linewidth}
        \includegraphics[width=\linewidth]{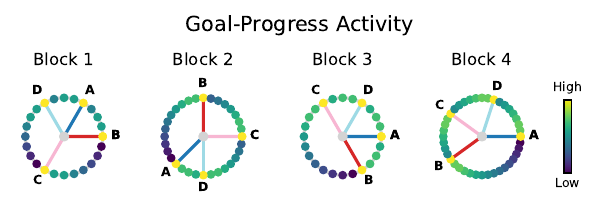}
        \subcaption[]{\label{fig:progress}}
    \end{minipage}
    \begin{minipage}[b]{0.49\linewidth}
        \includegraphics[width=\linewidth]{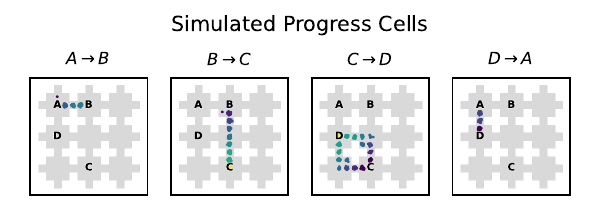}
        \subcaption[]{\label{fig:progress_activations}}
    \end{minipage}
        \begin{minipage}[b]{0.49\linewidth}
        \includegraphics[width=\linewidth]{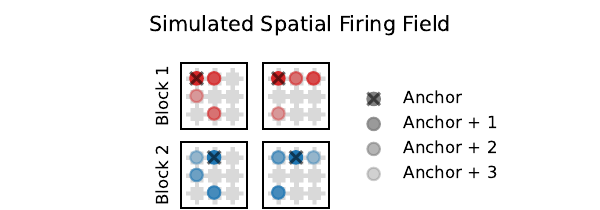}
        \subcaption[]{\label{fig:anchor}}
    
    \end{minipage}
    \begin{minipage}[b]{0.49\linewidth}
        \includegraphics[width=\linewidth]{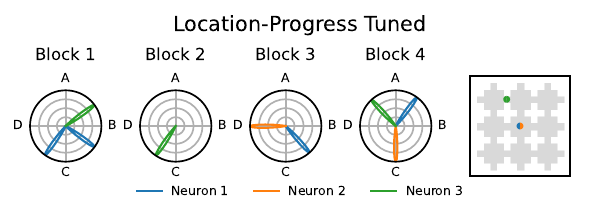}
        \subcaption[]{\label{fig:polar_activations}}
    \end{minipage}
\begin{minipage}[b]{0.49\linewidth}
        \includegraphics[width=\linewidth]{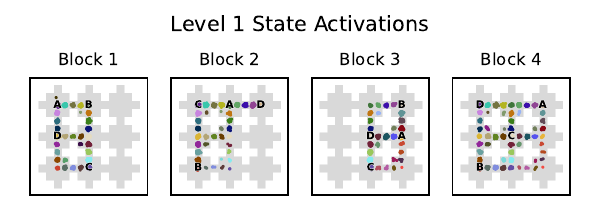}
        \subcaption[]{\label{fig:low_activations}}
    \end{minipage}
    \caption{\textbf{Simulation of neural activity in the S-HAI model during the ABCD task.} 
    \label{fig:neural_panel}
\small{(a) Four blocks of the ABCD task.  
(b) Simulated activity of ``goal-progress'' cells during the ABCD task, plotted in a circle. The nodes represent locations visited during the task and their colors nodes indicate the normalized expected inductive cost, under the variational posterior over the Level 1 state; these range from purple (low expectation, corresponding to early phase of progress towards a goal) to green (mid expectation and mid phase of progress) and yellow (high expectation and late phase of progress). The line indicates where the reward was received, and the color indicates which phase of the task this was in. 
(c) Simulated activity of ``goal-progress'' cells during the ABCD task, plotted in space. The panels show activations of the normalized expected inductive cost under the variational posterior over Level 1 state in distinct phases of Block 4. As before, purple, green, and yellow represent low expectation (early phase), mid expectation (mid phase), and high expectation (late phase of progress). Note that in the third panel, there are two paths from C to D, i.e., the animal can move upward or to the left after leaving C. 
(d) Simulated spatial firing fields of the grounding likelihood. Shaded nodes illustrate lagged fields for anchors placed at goal locations in Block 1 (Top) and the resulting remapping in Block 2 (Bottom). This architecture simulates neurons that fire at fixed task-space lags from an anchor. 
(e) Polar plot depicting simulated neural tuning of three neurons coding for the conjunction of spatial location and progress toward a goal. The right plot indicates the spatial tuning for each of the three neurons, shown in blue (Neuron 1), orange (Neuron 2), and green (Neuron 3). Neurons 1 and 3 are tuned to ``mid'' progress, and Neuron 2 to ``late'' progress. (f) Level 1 state activations during the last 1000 steps in each environment. Dots are plotted on the agent's tile and colored by the low-level state. Locations are augmented with random noise for visual clarity. 
See the main text for further explanation. See Table~\ref{tab:funccorr} for descriptions on how these neural correspondences are created}
}
\end{figure}

In this simulation, we aim to assess what internal representations emerge in the schema-based (S-HAI) agent during schema learning and how they relate to neural codes reported in the medial prefrontal cortex (mPFC) of rodents performing the ABCD task \citep{el2024cellular}. According to the Bayesian brain hypothesis, neurons do not simply fire in response to stimuli; rather, their activations encode probabilistic beliefs about relevant quantities in the environment \citep{doya2007bayesian,parr2022active}. To simulate neural activity, we therefore interpret neurons as representing beliefs over particular states, goals, or transitions. We focus our simulations on four ABCD problems, depicted in Figure~\ref{fig:tasks_neural}. This allows us to observe which neural beliefs evolve as the agent performs the problems, which remain invariant or vary across problem instances, and how they map to neural activation reported in the rodent mPFC.

A key finding of the ABCD study \citep{el2024cellular} is that the rodent mPFC encodes a large population of cells tuned to various combinations of goal-related, spatial, and other types of task-related information. Among these, the most frequent are "goal-progress cells," or cells that are tuned (mainly) to progress (e.g., early, mid, and late phases) toward abstract goals, independently of goal identity or physical distance. This is evidenced by the fact that the firing of these cells occurs when approaching any goal and stretches or shrinks depending on the spatial distribution of goal locations; see Figure 2c in \citep{el2024cellular}. Note that goal-progress cells have been reported also in other studies in the mPFC \citep{wang2026generalized} and in other brain areas, such as the nucleus accumbens (NAc)  \citep{takakuwa2026multiplexed}.

Neural activations that track progress toward goals naturally emerge in our model when considering the agent's belief over the inductive cost, under the expectation of its current location. During planning, the inductive cost associated with each state reflects how far this is from the preference state (see method for details). Figure~\ref{fig:progress} shows simulated neural activity in our model, where the expected inductive cost is normalized across consecutive steps, revealing that this value increases as the agent approaches the goal. Notably, the agent’s goal expectation ramps up consistently as it approaches the next goal, regardless of which goal is targeted. This is illustrated, for example, at the start of the sequence, its expectation is initially low (early progress, purple node), then increases as it moves toward goal C, passing through mid expectation (green node) before reaching high expectation (yellow node). This pattern corresponds to the engagement of distinct populations of “goal-progress cells”, tuned to early, mid, or late phases of progress toward any goal.


Simulated populations of “goal-progress cells” are further illustrated in Figure~\ref{fig:progress_activations}, where their activations are plotted on top of the maze layout during Block 1. The agent’s trajectories are shown as colored dots (with small amounts of noise added to avoid overlapping points). In the third panel, two distinct trajectories can be observed between C and D. The dots are color-coded as before, indicating that along each trajectory (e.g., from A to B in the first panel), different populations of “goal-progress cells” become active in sequence: first those tuned to mid goal expectation (green), then to high goal expectation (yellow), and finally to low goal expectation after the reward is collected at the goal location (purple). Importantly, this ramping pattern emerges consistently across different trajectories, independent of the specific goal destination.
In this simple implementation, goal-progress cells reduce to goal-distance cells, as progress is measured relative to the latent states at Level 1, which map directly onto physical locations. To ensure that the visualized values are comparable across blocks, we normalize the expected inductive cost values.

Another key finding of the ABCD study \citep{el2024cellular} is that neurons do not generalize their task-state preference (A, B, C, or D) across blocks; instead, they remap. However, this remapping is modular: within each module (or “buffer”), one neuron serves as an ``anchor'' tuned to a specific conjunction of goal progress and location, while other neurons fire at fixed lags in task space relative to that anchor. In this way, the neurons belonging to the same module conserved their within-module tuning relations across tasks.

In our model, the grounding likelihood provides a natural scaffold for the modular remapping of goal codes in the context of planning. Under the active inference framework, actions are selected by scoring the imagined future using the expected free energy. This rollout can be considered “egocentric”—centered on the agent’s current location (e.g., the location of goal A immediately after it has been collected) within the ABCD plan—such that the planning tree encodes multiple expected positions of the agent in task space after n steps (goal B at the next step, goal C after two steps, and so on), all in parallel \citep{donnarumma2026inferential}. Figure~\ref{fig:anchor} illustrates the simulated neural firing field of the selected high level goal locations, remapped through the grounding likelihood tensor. In each plot, the anchor corresponds to a specific goal location (marked with an X), and the remaining cells encode future goal locations as task-space offsets from this anchor, with increasingly pale shading indicating greater offset. 

The first column of Figure~\ref{fig:anchor} illustrates this when location 1 (marked X) is anchored: in Block 1, this location corresponds to goal A, and the module encodes A (anchor), B (anchor+1), C (anchor+2), and D (anchor+3), shown in progressively lighter shades of red. In Block 2, the same location corresponds to goal C, and the module correspondingly remaps to encode C (anchor), D (anchor+1), A (anchor+2), and B (anchor+3). The second column show analogous modular remapping when the anchor is located in position 2. These examples illustrate that while individual neurons remap and change their goal preference across blocks, the internal task-space structure of the module is preserved, consistent with empirical findings and the "buffers" model of \citep{el2024cellular}. 

The study of \citep{el2024cellular} further reports that modular remapping is not restricted to goal locations but extends to anchors at intermediate locations between goals, in conjunction with specific, fine-grained goal-progress phases. In our model, this phenomenon can be simulated by anchoring the activity of “goal-progress cells” (shown in Figure~\ref{fig:progress}) to specific spatial locations. Figure~\ref{fig:polar_activations} illustrates this effect. Here, the activity of goal-progress neurons (discretized into “early,” “mid,” and “late” progress bins) is projected onto a task-centric polar plot. Cardinal angles (0°, 90°, 180°, 270°) represent the four goals of the ABCD loop, while colored neurons are anchored to the spatial locations shown in the adjacent plot. Neuron 1 (blue) and Neuron 2 (orange) are anchored to the same spatial location but encode different progress phases (mid vs. late), whereas Neuron 3 (green) shares mid-progress tuning with Neuron 1 but is anchored to a different location. In this example, neurons appear to remap their task-state preference as goal locations change across Blocks 1–4. However, the anchors—specific conjunctions of progress phase and spatial location—remain invariant across tasks, consistent with the empirical findings of \citep{el2024cellular}. For example, the apparent remapping of Neuron 3—from the $A \rightarrow B$ transition in Block 1 to the $C \rightarrow D$ transition in Block 4—reflects the fact that its anchor location corresponds to mid goal-progress between $A$ and $B$ in Block 1, but to mid goal-progress between $C$ and $D$ in Block 4. This result shows that internal task variables—specifically, goal-progress representations—provide a principled computational basis for the modular remapping observed in the study of \citep{el2024cellular}.


Finally, beyond the neural codes that support navigation in task space at Level 2—potentially associated with frontal cortical mechanisms—our model also is consistent with for neural codes that support navigation in physical space at Level 1, possibly linked to spatial mapping and navigation in the hippocampal formation \citep{nyberg2022spatial}. Figure~\ref{fig:low_activations} shows Level 1 spatial activations, which encode the agent’s position within the environment. Regardless of the task instance, each spatial location corresponds to a distinct activation pattern. For example, the bottom-right location consistently activates the same neuron, analogous to hippocampal place cells \citep{o1971hippocampus}. This suggests that Level 1 neurons provide a stable spatial representation independent of the current task, enabling the model to preserve consistent positional information while Level 2 representations capture task-specific and goal-directed information.


\section{Discussion}

The traditional view of learning in psychology, neuroscience, and AI emphasizes the gradual accumulation of experience. Alongside classical learning theories, it has long been postulated that humans and other animals are capable of forming \emph{schemas}—data structures that encode structural relations between events while abstracting away from sensory details—and of rapidly reusing them to generalize knowledge to novel contexts by quickly rebinding new experiences to existing schemas \citep{piaget1952origins,bartlett1932remembering}. A growing body of literature, recently reviewed in \citep{farzanfar2023cognitive}, supports the idea that advanced cognitive abilities, such as fast generalization and the abstraction of knowledge across contexts, depend on schema-based mechanisms possibly involving the hippocampus, entorhinal cortex, and frontal cortex.

This study introduces a novel computational approach—schema-based hierarchical active inference (S-HAI)—that addresses the formation of schemas from experience and their rapid generalization to novel contexts. The model builds on theories of hierarchical predictive processing and active inference \citep{parr2022active,VandeMaele2024,pezzulo2018hierarchical,butz2025contextualizing,smith2022step,lanillos2021activeinferenceroboticsartificial,matsumoto2020,friston2021,taniguchi2022,isomura2018} and extends them with schema-based mechanisms. The S-HAI agent is organized hierarchically: the higher level (Level 2) is responsible for schema learning and navigation in an abstract task space, while the lower level (Level 1) encodes spatial information and supports navigation in physical space. Crucially, the two levels are connected by a mechanism unique to our model, the \emph{grounding likelihood}, which maps abstract goal codes in the schema to physical locations. The rapid learning of this mapping enables the agent to flexibly generalize the same schema to novel goal configurations.

With a series of simulations, we validate the ability of the schema-based S-HAI agent to recapitulate behavioral findings in tasks requiring rapid generalization, as well as neural findings reported in the medial frontal cortex of rodents performing such tasks. Our results show that, after learning a schema for a general class of navigation problems sharing the same structure—namely, problems requiring cyclic visits to four goal locations to obtain rewards, as in the ABCD task \citep{el2024cellular}—the S-HAI agent exhibits rapid online generalization to novel problems in which the relational structure is preserved but the spatial locations of the four goals change. Remarkably, the schema-based agent performs on par with an agent without schemas trained offline on all problems, while also showing greater flexibility in adapting to variations of the task. 
Our simulations also show that the same approach can be generalized to more challenging tasks in which multiple goals can share the same location (ABCB task), analogous to spatial alternation tasks \citep{jadhav2012awake}, and that the S-HAI agent can learn online and select among a mixture of likelihood mappings between the abstract schema and concrete problems, demonstrating the ability to decide when to reuse an existing mapping or create a novel one—capturing at least basic signatures of the assimilation of novel experiences into existing relational structures \citep{piaget1952origins}. Finally, and importantly, the S-HAI model generates patterns analogous to key neural signatures of schemas identified in the medial frontal cortex of rodents performing the ABCD task \citep{el2024cellular}, most prominently capturing the activity of goal-progress cells, while also reflecting the heterogeneous coding of other cells sensitive to combinations of goal, spatial, and task-related information.

Taken together, these results support S-HAI as a candidate computational framework that demonstrates the efficacy of schema-based learning and inference, capturing both behavioral and neural signatures of rapid generalization, flexible problem solving, and the assimilation of new information into existing schemas. Schema processing has been linked to at least four subprocesses, including \emph{schema acquisition} (the process by which a schema is abstracted from the relational structure shared across multiple experiences), \emph{schema representation} (the format in which this abstracted structure is encoded and related to low-level sensory details), \emph{schema-guided inference} (the use of an existing schema to fill in missing information and predict outcomes within a familiar situation), schema-based exploration (the use of a schema to speed up exploration in novel domains), and \emph{schema-guided generalization (or transfer)} (the application of an existing schema to a novel situation that shares its relational structure but differs in surface features) \citep{piaget1952origins,bartlett1932remembering,gilboa2017neurobiology,tse2007schemas,bein2025schemas}. The S-HAI model speaks, at least partially, to each of these subprocesses. It provides a candidate mechanism through which schemas and relational structures can be abstracted from specific experiences, with the caveat that, by design, our model does not capture the more gradual process through which animals may extract schematic representations, which is often linked to systems consolidation \citep{dudai2012restless}. It also suggests a possible format—based on CSCGs and grounding likelihoods—for representing schemas and connecting them to specific experiences. Furthermore, it proposes a specific inferential mechanism through which schemas can be used to predict the next step (in our simulations, the next navigation goal). Furthermore, it highlights how a schema can accelerate exploration; for example, in a new ABCD task, an agent can stop searching for novel goals once it finds the D goal. Finally, it highlights the importance of learning the grounding likelihood as a mechanism for generalizing and transferring schemas to novel situations. Taken together, these components illustrate how schema acquisition, representation, inference, and transfer can be addressed within a common computational framework, grounded in principles of predictive processing and hierarchical active inference \citep{parr2022active}. This unification within a predictive processing framework suggests that the same principles used to model perception, action, and decision-making may also contribute to schema formation, flexible reuse of relational knowledge, and generalization in the brain. 

By providing a mechanistic model of schema-based learning and inference, our framework also generates novel empirical predictions that can be tested in future experiments. One such prediction concerns the behavior and neural representations that might be observed in animals performing the ABCB task (Figure~\ref{fig:abcb_reward}). Our simulations suggest that correctly solving this task requires a mechanism (clone-based or similar) capable of disambiguating between distinct instances of the same goal (e.g., goal B). This, in turn, should produce specific patterns of behavior and Level 2 neural representations, including separate representations of the same goal when it is encountered twice (see Figure~\ref{fig:task3}). Another critical prediction concerns the functional role of the neural activity patterns reported in the medial frontal cortex of rodents during the ABCD task \citep{el2024cellular}. Our simulations further suggest that distinct components of the model—Level 2 beliefs about goals, Level 1 beliefs about locations, and the grounding likelihood—map onto distinct neuronal populations tuned to goal progress, spatial location, and their conjunctions. Consequently, perturbing these populations should produce dissociable behavioral effects. Disrupting neurons that encode beliefs about the current task phase should impair the animal’s ability to correctly infer its next goal. Disrupting neurons that encode the grounding likelihood should impair systematic remapping across blocks. Finally, disrupting neurons that encode beliefs about spatial location should impair path planning and navigation. These predictions remain to be tested in future experiments.

The capability of our model to flexibly rebind stimuli to existing structures and to adapt to novel tasks (with shared structure) on the fly, as opposed to doing so through the slow and gradual modification of model parameters, as is most commonly done in neural network research, is shared with other approaches. This characteristic is related to \emph{in-context} learning in transformer-based large language models, or the ability to generalize rapidly from a few examples of a new concept not present in their training set, without updating the model itself \citep{brown2020language,swaminathan2023schema}. It is also related to so-called \emph{meta-learning} and \emph{learning-to-learn} in recurrent neural networks, which learn generalizable patterns across diverse sequences and sets of stimuli and actions, allowing for new experiences to be rapidly incorporated within their recurrent dynamics \citep{goudar2023schema,binz2024meta,wang2021meta,wang2018prefrontal}. A similar principle is present in computational models such as the Tolman-Eichenbaum Machine, which separates a relational structure common across problems from problem-specific sensory representations that can be remapped to the same structure when the problem changes \citep{whittington2020tolman}. Related capabilities are also found in neural networks capable of rapid associative indexing or one-shot learning through fast learning rules \citep{yang2025episodic,wu2025simple,hwu2020neural,d2026genesis}. While these approaches differ extensively, they provide viable models of schema-related capabilities and the generalization of existing knowledge and skills to novel contexts. Assessing their relative strengths and weaknesses in relation to the predictive processing approach advanced here remains an open question for future research.

The current S-HAI agent has several limitations that can be addressed in future research. First, while our model considers multiple grounding likelihoods, it currently implements only a single schema. The S-HAI framework allows a straightforward extension from maintaining a mixture of grounding likelihoods to also maintaining a mixture over multiple schemas, thereby providing a more comprehensive account of the assimilation and accommodation processes envisaged by \citep{piaget1952origins}. 

Second, the model focuses primarily on schema learning and inference in the frontal cortex, based on probabilistic generative models that extend Hidden Markov Models (HMMs). This approach permits reproducing some key aspects of the neural coding of schemas in the frontal cortex, but cannot capture its full complexity. For example, although Level 2 representations in our model encode goals explicitly, neurons that generalize goal coding across blocks were not observed in the mPFC of the rodent study~\citep{el2024cellular}. Notably, however, goal-coding neurons that generalize across tasks have been reported in rhesus monkeys~\citep{Genovesio2012}. Whether a comparable form of abstract, task-general goal coding exists in the rodent brain remains an open empirical question. We also acknowledge that our experimental comparisons are qualitative; direct quantitative comparisons of learning dynamics and neural tuning with recorded data would provide more evidence of our model and remains an important goal for future work. 

Third, future studies could explore biologically realistic implementations of HMMs \citep{kappel2014stdp} and more systematically investigate the mapping between these models, the ``structured memory buffer'' model of \citep{el2024cellular}, and neural computations in the frontal cortex. Additionally, future work might extend S-HAI to provide a system-level model addressing the compositional construction of behavioral sequences \citep{manakov2026universal} and schema-based processes beyond the frontal cortex, encompassing other relevant brain regions such as the hippocampus and entorhinal cortex. More broadly, future extensions of the model could consider a more comprehensive neural account of schema-based processing, encompassing additional aspects such as the role of schemas in rapid systems consolidation and cortical memory formation, individual variability in schema acquisition,
interactions between episodic and semantic memory in schema formation and processing, and the role of hippocampal replay in these processes \citep{lewis2011overlapping,bhattarai2026offline}, which are not considered in the present model.

Fourth, as noted above, the model implements a single schema and therefore does not address Piaget's \emph{accommodation}, which involves creating new schemas or modifying existing ones when new information cannot be assimilated into existing schemas. By the same token, the model cannot address the competition between multiple schemas, consistent or inconsistent with the current situation. The formation of novel schemas could be implemented with a Dirichlet process over Level 2 transition models, analogous to the mixture of grounding likelihoods, whereby information inconsistent with existing schemas (i.e., exceeding a surprise threshold) triggers the formation of a new schema rather than assimilation into an existing one. A more sophisticated architecture, endowed with modules and gating functions, could also enable compositionality and the recombination of multiple schemas \citep{bakermans2025compositional,riveland2026compositionality}. Such architecture would potentially allow for extending existing schemas, e.g., from ABCD to ABCDE. These limitations can be addressed in future work.

Fifth, in this study we examined schema formation in S-HAI using variants of the ABCD spatial navigation paradigm. Schemas, however, have been studied across a much broader range of memory domains, including associative, semantic, contextual, and episodic memory (e.g., \citep{tse2007schemas,gilboa2017neurobiology,takeuchi2022brain,audrain2022schemas,bhattarai2026offline,thorndyke1979use}). Crucially, none of the mechanisms used here relies on assumptions specific to spatial navigation. The CSCG can process any type of sequential data and has been used to model sequential structures beyond navigation, such as letter sequences \citep{swaminathan2023schema}. Similarly, the grounding likelihood makes no assumptions specific to spatial navigation; it simply encodes a state-to-state mapping. Therefore, we expect the core mechanisms of S-HAI to extend to spatial and non-spatial problems in which a shared relational structure must be mapped onto new sets of observations.

Finally, future work could investigate how schema-based mechanisms can be reused to support navigation in abstract, conceptual spaces. Recent studies suggest that the brain may rely on shared computational mechanisms for mapping and navigation in both physical and conceptual domains, with a central role played by the hippocampal–entorhinal system \citep{buzsaki2013memory,bellmund2018navigating,vigano2023mental,bottini2020knowledge,dong2024grid}. Understanding how schema formation and schema-based inference contribute to constructing and navigating such abstract cognitive maps could provide a unifying framework for explaining flexible cognition across spatial and non-spatial domains.



\section{Methods}
\label{sect:methods}

    Our approach builds on active inference, a framework in which agents minimize variational free energy by updating beliefs (perception), choosing actions (policy evaluation), and adapting model parameters (learning)~\citep{parr2022active,smith2022step}. We expand the paradigm by introducing a generative model that can represent and reuse abstract \emph{schemas} -- structured, generalizable representations of task dynamics -- across multiple environments. Instead of learning the specific sequence of rewarded locations in a task~\citep{VandeMaele2024}, the schema captures the abstract structure of this task, such as the fact that there are four rewards in four different locations (as in the ABCD task of \citep{el2024cellular}) or three rewards in alternating locations (as in the spatial alternation task of \citep{jadhav2012awake}). In various task instances, this schema can then be mapped probabilistically to environment-specific states. This corresponds to a very fast learning process, since the agent only needs to learn a new mapping (which we call a \emph{grounding likelihood}) from the abstract schema states to environment-specific locations. The use of schemas, therefore, allows the agent to rapidly generalize and transfer high-level knowledge across different task instances. 

    In this section, we first briefly review the functioning of active inference, and then we illustrate the structure of the novel schema-based (S-HAI) agent.
    
    
     \subsection{Active Inference}
     \label{sect:aif}

    Active inference is a framework that describes cognitive processes and brain dynamics in living organisms, in terms of the minimization of an information-theoretic functional: variational free energy ~\citep{parr2022active}. active inference agents are endowed with a \emph{generative model}: a probabilistic model that encodes the internal belief over the causal relationship between hidden states, actions, and subsequent observed outcomes. Note that this is distinct from the true physical process (called \emph{generative process}) that generates the outcomes in the world. 

    As agents are computationally bound, inference over the posterior becomes intractable for large state spaces. Therefore, agents use approximate (variational) inference by minimizing their variational free energy, a bound on the surprise, defined as: 
    
    \begin{align}
        \mathcal{F} &= \mathbb{E}_{Q(\mathbf{\tilde{s}};\theta)}[\ln Q(\mathbf{\tilde{s}};\theta) - \ln P(\mathbf{\tilde{s}}, \mathbf{\tilde{o}}, \mathbf{\tilde{a}})] \\
        &= \text{D}_\text{kl} [Q(\mathbf{\tilde{s};\theta}) || P(\mathbf{\tilde{s}}| \mathbf{\tilde{o}}, \mathbf{\tilde{a}})] - \ln P(\mathbf{\tilde{o}}), 
    \end{align}
    
    with variational posterior $Q(\mathbf{\tilde{s}}; \theta)$, parameterized by $\theta$ over the latent state trajectories. When $\mathcal{F}$ is minimized, the variational posterior approximates the true posterior. 


    A key assumption of active inference is that organisms minimize their variational free energy by changing their beliefs about the world to better fit their observations (perception), by changing the parameters of their generative model (learning), or by changing the world -- and therefore their observations -- by acting (action selection and planning).

    During perception, the generative model is inverted to find the hidden causes that give rise to the observed outcomes. this is done by minimizing the free energy $\mathcal{F}$ with respect to the approximate posterior's variational parameters $\theta$. by applying a mean field factorization over time $Q(\mathbf{\tilde{s}};\theta) = \prod_t Q(\mathbf{s}_t;\theta_t)$~\citep{beal_variational_nodate} and parameterizing the approximate posterior at each timestep as a categorical distribution $Q(\mathbf{s}_t) = \text{Cat}(\theta_t)$, it is possible to perform approximate posterior inference over hidden states, using standard mean-field variational updates~\citep{pymdp}.


    During learning, the agent's generative model is updated through its experiences. This is achieved by optimizing the variational free energy objective with respect to the model parameters. Both the observation and dynamics likelihood are modeled as a categorical distribution, with a conjugate Dirichlet prior over the model parameters. The variational posterior distribution of the Categorical-Dirichlet pair is the sum of the sufficient statistics of the observed data with the prior~\citep{Blei_2017}. For the observation likelihood, the sufficient statistics are computed as the outer product between the observation and state at the current timestep $(\mathbf{s}_{t},\mathbf{a}_{t})$. While for the dynamics likelihood, the sufficient statistic is computed as the outer product between the state at the current time step, and the state-action pair at the previous timestep $(\mathbf{s}_{t},\mathbf{s}_{t-1},\mathbf{a}_{t-1})$. The belief over the state $\mathbf{s}_t$ is found using a Bayesian filtering approach when learning non-clone structured transition dynamics, and using smoothing when learning the dynamics of a CSCG online. In this work, we consider a fixed observation likelihood and only learn the parameters of the dynamics and the grounding likelihood (see Section~\ref{sect:grounding_likelihood}). 

    Finally, during action selection, the agent selects the policy $\pi$ (a sequence of actions) that minimizes the expected free energy $\mathcal{G}$, i.e., the expectation of the free energy under future observations, under each policy, augmented with an inductive prior over policies $\mathcal{H}$, which enables intended behavior towards preferred states, rather than preferred outcomes~\citep{friston2023active}. 
    
    The expected free energy over a policy is the sum of the expected free energies at each future timestep $\tau$: $\mathcal{G}(\pi) = \sum_\tau \mathcal{G}_\tau(\pi_\tau)$. 
    \begin{align}
        \mathcal{G}_\tau(\pi_\tau) &= \mathbb{E}_{Q(\mathbf{o}_\tau |\mathbf{\pi_\tau})}\big[\ln Q(\mathbf{s}_\tau| \pi) - \ln P(\mathbf{o}_\tau, \mathbf{s}_\tau|\pi_\tau) \big] \\ 
        &= - \underbrace{\mathbb{E}_{Q(\mathbf{o}_\tau|\pi_\tau)} \bigg[ \text{D}_\text{KL}\big[ Q(\mathbf{s}_\tau, \mathbf{o}_\tau | \pi_\tau) || Q(\mathbf{s}_\tau | \pi_\tau) \big]  \bigg]}_{\text{Epistemic Value}} - \underbrace{\mathbb{E}_{Q(\mathbf{o}_\tau|\pi_\tau)}\bigg[\ln P(\mathbf{o}_\tau)\bigg]}_\text{Utility}
    \end{align}
    
    This quantity trades off \emph{epistemic value}, scoring how much information is gained from pursuing the policy and \emph{utility}, scoring how much the policy realizes the agent's goals in the future, encoded as prior preferences over observations $P(\mathbf{o}_\tau$). Similar to the expected free energy, the inductive prior over policies is simply the sum of the inductive priors at each future timestep $\tau$: $\mathcal{H}(\pi) = \sum_\tau \mathcal{H}_\tau(\pi_\tau)$.

    \begin{equation}
        \mathcal{H}_\tau(\pi_\tau) = - \underbrace{\mathbb{E}_{q(\mathbf{s}_\tau | \pi_\tau)}[ \ln P(\mathbf{s}_\tau)]}_\text{Inductive cost}
        \label{eq:inductivecost}
    \end{equation}

     This quantity scores future states $\mathbf{s}_\tau$ with respect to an intended future state, encoded as a prior over future state $P(\mathbf{s}_\tau)$. Crucially, unlike the utility term (conditioned on an expected outcome), the \textit{inductive cost} propagates through the latent dynamics across the state space of the agent as it assigns a cost value to each state, reflecting the distance to the intended (goal) state. For a more comprehensive description, the reader is referred to~\citep{friston2023active}.  
     
    The approximate posterior over the policies $Q(\pi)$ is set to be proportional to the expected free energy and inductive prior: $Q(\pi) = \sigma(-\gamma (\mathcal{G}(\pi) + \mathcal{H}(\pi)))$. Here, $\sigma$ is the softmax function, and $\gamma$ is the sampling temperature, where lower temperatures yield more stochastic behavior. This implies that the more a policy is expected to minimize free energy, the higher the probability of selecting it. For a more detailed explanation of active inference, see \citep{parr2022active}.
        
    \subsection{Formal description of the schema-based hierarchical active inference agent}

    \begin{figure}[t!]
        \centering
        \includegraphics[width=\textwidth]{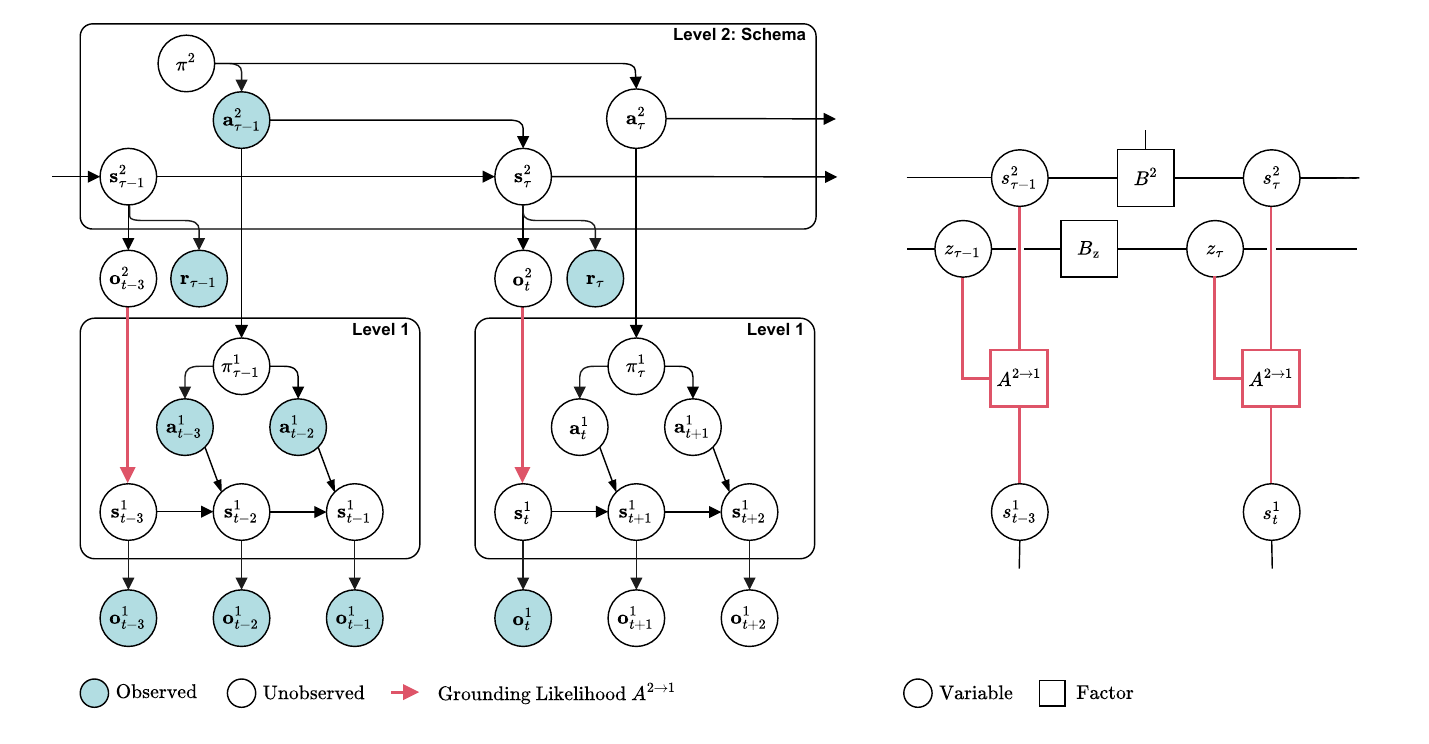}
        \begin{minipage}[b]{0.5\linewidth}
            \subcaption[]{\label{fig:bayesnet}}
        \end{minipage}
        \begin{minipage}[b]{0.45\linewidth}
            \subcaption[]{\label{fig:factorgraph}}
        \end{minipage}
        \caption{\textbf{The hierarchical generative model} \small{(a) The hierarchical generative model used in this study. The observed variables are denoted in green, while unobserved variables are denoted in white. The red arrow indicates the \emph{grounding likelihood}. Note that the policy of Level 1 is conditioned on the high-level action with a corresponding preferred Level 2 observation, which maps to a Level 1 goal (i.e. a preferred Level 1 state) through the same grounding likelihood (e.g., if the high level action is ``move to B'', then the low level preference would be grounded to level 1 state 192 as shown in Figure~\ref{fig:hierarchicalmodel}).} (b) A factor graph representation of the grounding likelihood that maps states at Level 2 to states at Level 1. $B_z$ shows the transition dynamics in a mixture of grounding likelihoods. The red connections refer to the same likelihood as in Figure~\ref{fig:bayesnet}. See the main text for explanation.}
    \end{figure}

Active inference agents are endowed with a generative model, which essentially defines and constrains their knowledge and capabilities. In this study, we propose a novel type of hierarchical generative model that allows the agent to engage in schema-based learning and inference, as seen for example in the ABCD task of \citep{el2024cellular}, where the agent navigates in space to reach sequences of four goals.

Figure~\ref{fig:bayesnet} shows the generative model for schema-based hierarchical active inference (S-HAI), using the formalism of Bayesian networks (see also Figure \ref{fig:hierarchicalmodel} for a more informal schematization). It comprises two hierarchical layers. The bottom layer (Level 1) operates at the finest timescale and handles the agent's spatial localization and navigation. At this level, the agent receives direct observations of its location and acts through movement. Top-down goals are set as preferences over future states at the lower level that the agent needs to reach. 

The higher layer (Level 2) implements schema-based inference and learning. The schema operates at a slower timescale compared to Level 1 and captures the abstract task structure, integrating information about reward and bottom-up messages containing the inferred state at the lower level. The dynamics at this level model the state transitions across the goal states that provide reward observations~\citep{friston2024pixelsplanningscalefreeactive}. For example, in the ABCD task, the schema captures an abstract sequence of actions to move to the next goal, cyclically (from A to B, C, D, and then again to A, etc.).

The hierarchical generative model comprises two coupled Partially Observable Markov Decision Processes (POMDPs), one per level, which interact through top-down and bottom-up message passing~\citep{VandeMaele2024,catal2021}. Each layer maintains its generative model, and for the generic level~$i$, the joint distribution factorizes as:
    \begin{equation}
        P(\mathbf{\tilde{s}}^i, \mathbf{\tilde{o}}^i, \mathbf{\tilde{a}}^i ) = P(\mathbf{s}^i_0) \prod_{t=1}^T P(\mathbf{o}^i_t | \mathbf{s}^i_t ) P(\mathbf{s}^i_{t} | \mathbf{s}^i_{t-1}, \mathbf{a}^i_{t-1}) P(\mathbf{a}^i_{t-1})
    \end{equation}
    with states $\mathbf{\tilde{s}}^i$, actions $\mathbf{\tilde{a}}^i$, and observations $\mathbf{\tilde{o}}^i$. Note that the tilde indicates a sequence of that variable over time. The superscript indicates the hierarchical level, and the subscript indicates time. 
    
    The POMDP at each level factorizes into a prior $P(\mathbf{s}^i_0)$ over the initial state, a prior $P(\mathbf{a}^i_{t-1})$ over action, an observation likelihood $P(\mathbf{o}^i_t | \mathbf{s}^i_t)$ that maps the current state to an observation, and a dynamics model $P(\mathbf{s}^1_{t+1} |  \mathbf{s}^1_{t-1}, \mathbf{a}^1_{t-1})$ that maps the current state-action pair to future state.

The two POMDPs continuously engage in hierarchical message passing to realize hierarchical active inference. Inferred states of Level 1 are passed as bottom-up messages to the observations for Level 2, while top-down actions from Level 2 specify goals at Level 1 (see Figure~\ref{fig:bayesnet}). 

Crucially, to support schema-based inference, the agent cannot assume a one-to-one mapping between the states at Level 1 $\mathbf{s}^1_t$ and the observations at Level 2 $\mathbf{o}^2_t$. For example, the agent cannot assume that in the ABCD task, the A goal is always in the same spatial position, as it varies across trials. For this, the agent must learn a \textit{grounding likelihood} $P(\mathbf{s}^1_t|\mathbf{o}^2_t)$ that maps from observations at Level 2 to states at Level 1 (e.g., it tells the agent in which spatial position the goal A is), represented by the red arrow in Figure~\ref{fig:bayesnet}. More details on the \textit{grounding likelihood} are provided in Section  \ref{sect:grounding_likelihood}.

For the first level, the observations $\mathbf{o}^1$ are the color of the tile the agent is currently occupying (see the ABCD Task in Figure ~\ref{fig:taskABCD}). The agent can perform primary movement actions $\mathbf{a}^1$, i.e., move up, down, left, or right. Through a clone-structured likelihood mapping, an observation maps to a uniform distribution over ``clone'' states. Through the encountered transitions between states, the agent infers the Level 1 state $\mathbf{s}^1$. At the higher level, the observations are received bottom-up using the grounding likelihood, on rewarding timesteps, and the action space consists of only one action: ``move to the next goal in the sequence''. Gating bottom-up messages based on reward is a form of active data selection~\citep{friston2024pixelsplanningscalefreeactive}, ensuring that the high-level learns the dynamics of the rewarding states at a coarser timescale. 
    
    Through this hierarchical coupling, the grounding likelihood mediates how the abstract schema at Level 2 specifies concrete goals at Level 1, allowing the agent to generalize task structure across environments.

    \subsubsection{The grounding likelihood}
    \label{sect:grounding_likelihood}
    
    The grounding likelihood acts as the interface between the high-level schema that tracks the abstract task states and the lower-level state, which is responsible for spatial navigation. By translating schema states to specific spatial states, the same abstract task schema can be reused and generalized across multiple instances of the same abstract task, even though they differ in their specifics. In particular, the likelihood describes the mapping of a schema ``observation'' $\mathbf{o}^2_t$ to a spatial state $\mathbf{s}^1_t$. This schema observation is decoupled from direct reward signals (see the red arrow in Figure~\ref{fig:bayesnet}), and thus encodes an ``abstract spatial state'' within the schema space.  Note that when the grounding likelihood is the identity mapping, this task space observation has to learn the specific location of rewards and their dynamics for each of the tasks~\citep{VandeMaele2024}. This prevents generalization of the schema abstraction to other task instances. 
        
    Formally, we model the grounding likelihood as a categorical distribution, conditioned on the high-level state indicating in which phase of the task the agent is. This likelihood specifies which spatial state the agent expects to find rewards: 
    \begin{equation}
        P(\mathbf{s}^1_t | \mathbf{o}^2_t) = \text{Cat}(A^{2\rightarrow1}_{\mathbf{o}^2_t}),
    \end{equation}

    Where the subscript $\mathbf{o}^2_t$ indicates slicing the parameter tensor $A^{2 \rightarrow 1}$ under the high-level observation. This distribution has a Dirichlet prior over the parameters with pseudo-counts $\alpha$. The variational posterior over these parameters is computed online using conjugate updates~\citep{Friston2016}. 

    Remember from Section~\ref{sect:methods} that the schema level only learns the dynamics of the rewarding sequence. Therefore, for updating the model, we use a modified belief vector $\hat{\mathbf{s}}^1_t$ to encode the probability of observing a reward at each state. In case the reward is observed, this corresponds exactly to the lower level state $\mathbf{s}^1_t$; However, in case the reward is not observed, we set this belief to a uniform distribution over all other locations, and a zero probability to the currently observed one. 
    \begin{equation}
        \hat{\mathbf{s}}^1_t = 
        \begin{cases}
         \mathbf{s}^1_t & \text{ if } r_t = 1, \\
         \frac{1}{n-1}(1-\mathbf{s}^1_t) & \text{ if } r_t = 0, 
        \end{cases}
    \end{equation}
    where $r_t$ is the observed reward at time t, $n$ is the number of low-level states, and $(1-\mathbf{s}^1_t)$ is a vector of all states except the current one, normalized by $n-1$ to ensure a proper probability distribution.

    With this definition of modified belief vector $\hat{\mathbf{s}}^1_t$, the variational posterior is updated using the update rule as derived in~\citep{pymdp}, and described in Section~\ref{sect:aif}:
    
    \vspace{0.5em}
    \begin{minipage}{0.45\textwidth}
        \begin{equation}
            P(A^{2 \rightarrow 1}) \sim \text{Dir}(\alpha)
        \end{equation}
    \end{minipage}   
    \hfill
    \begin{minipage}{0.45\textwidth}
        \begin{equation}
            \alpha_{t} = \alpha_{t-1} + \eta ( \hat{\mathbf{s}}^1_t \otimes \mathbf{o}^2_t) ,
        \end{equation}
    \end{minipage}
    \vspace{0.5em}
    
    Where $\mathbf{o}^2_t$ is a one-hot representation of the high-level observation, i.e., the abstraction over the location of the current reward in the task. $\eta$ is the learning rate, which is 1 in case of observed reward or $0.05$ otherwise, ensuring that the grounding likelihood is strongly updated by rewarding states and weakly by non-rewarding ones. Finally, $\otimes$ denotes the outer product, which increments the pseudocounts $\alpha$ on the co-occurrence of high-level observation and the low-level state. In the case that the rewarding state has been found, this will enforce the mapping of high-level observation to the low-level state, and in the other case, it will decrease the strength of the current low-level state by incrementing all other low-level states. 

    The Level 1 goal ($\mathbf{s}_\tau^1$ in Eq~\eqref{eq:inductivecost}) is specified by reusing the grounding likelihood in the top-down direction. Writing $\alpha_{\mathbf{o}^2_\tau}$ for the concentrations it has accumulated for outcome $\mathbf{o}^2_\tau$, we read the likelihood out with precision $\gamma$ using the softmax $\sigma$,
    $P_\gamma(\mathbf{s}^1_\tau | \mathbf{o}^2_\tau) = \sigma\!\left(\gamma \, \alpha_{\mathbf{o}^2_\tau}\right)$. Under the Level 2 action $\mathbf{a}^2_{\tau-1}$ and the current Level 2 state $\mathbf{s}^2_{\tau-1}$, the preferred Level 1 state distribution follows by marginalizing over the expected Level 2 state and outcome:
    \begin{align}
        P(\mathbf{s}_\tau^1 | \mathbf{a}^2_{\tau-1}, \mathbf{s}^2_{\tau-1}) =
        \sum_{\mathbf{o}^2_\tau} P_\gamma(\mathbf{s}_\tau^1 | \mathbf{o}_\tau^2)
        \sum_{\mathbf{s}^2_\tau} P(\mathbf{o}^2_\tau | \mathbf{s}_\tau^2)
        P(\mathbf{s}_\tau^2 | \mathbf{s}_{\tau-1}^2, \mathbf{a}_{\tau-1}^2),
    \end{align}
    from which the Level 1 goal is sampled,
    $\mathbf{s}^1_\tau \sim P(\mathbf{s}_\tau^1 | \mathbf{a}^2_{\tau-1}, \mathbf{s}^2_{\tau-1})$. This goal is only resampled when the lower level reaches it or encounters reward. 

    \subsubsection{Mixture of Grounding Likelihoods}

    A single grounding likelihood can be brittle when multiple tasks or environments are present. To support richer generalization and knowledge retention, we introduce a \textit{Mixture of Grounding Likelihoods} (MoGL), in which the agent maintains a set of grounding likelihoods and infers which one is active at each timestep.

    Formally, the low-level state is sampled from a mixture of grounding likelihoods:
    \begin{equation}
        P(\mathbf{s}_t^1 | \mathbf{o}_t^2) = \sum_{z_t} P(\mathbf{s}_t^1 | \mathbf{o}_t^2, z_t) P(z_t | z_{t-1}), 
    \end{equation}
    where $z_t$ is a discrete switching variable indicating which grounding likelihood is currently in use. We model $z_t$ as a Markov chain with fixed parameters, favoring staying in the same grounding likelihood with high probability and transitioning to a different grounding likelihood with low probability. See Figure~\ref{fig:factorgraph} for a factor graph visualization of the dynamics on switching variable $z_t$.

    The agent has to infer which grounding likelihood $z_t$ is currently in play. Similar to the inference mechanism described in Section~\ref{sect:aif}, we infer this belief over $z_t$, proportional to the expected log likelihood under the Dirichlet parameters of the selected grounding likelihood. In a similar vein to the modified belief in Section~\ref{sect:grounding_likelihood}, we invert the parameters of the grounding likelihood if it is a non-rewarding location: 
    
    \begin{equation}
        \hat{A}^{2 \rightarrow 1}_{z_t,\mathbf{o}^2_t} = 
        \begin{cases}
         A^{2 \rightarrow 1}_{z_t,\mathbf{o}^2_t} & \text{ if } r_t = 1, \\
         Z \cdot (1-\frac{1}{Z} A^{2 \rightarrow 1}_{z_t,\mathbf{o}^2_t}) & \text{ if } r_t = 0, 
        \end{cases}
    \end{equation}
    where $A^{2 \rightarrow 1}_{z_t,\mathbf{o}^2_t}$ is the vector of the concentration parameters for the Dirichlet distribution, conditioned on high-level observation $\mathbf{o}^2_t$, and switching variable $z_t$. Here, $Z$ is the sum of the concentration parameters. We infer the likelihood message as proportional to the expected log likelihood and combine it with the empirical prior $P(\mathbf{z}_t | \mathbf{z}_{t-1})$ from the Markov chain to find: 
    \begin{equation}
        Q(\mathbf{z}_t) \propto \sigma \big( \underbrace{\mathbb{E}_{P(\hat{A}^{2 \rightarrow 1}_{\mathbf{z}_t,\mathbf{o}^2_t})}[ \ln P(\mathbf{s}^1_t | \mathbf{o}^2_t; \hat{A}^{2 \rightarrow 1}_{\mathbf{z}_t,\mathbf{o}^2_t})}_{\text{Expected Log Likelihood}} ] \big) P(\mathbf{z}_t | \mathbf{z}_{t-1}),
        \label{eq:qz}
    \end{equation}
    
    Crucially, we model this mixture model as a non-parametric model that can expand the mixture by adding clusters~\citep{Stoianov2022,heins2025axiomlearningplaygames}, i.e., particular grounding likelihoods for observed maps. Formally, this is modeled as a truncated stick-breaking prior that expands the mixture model if the expected log likelihood, marked in Equation~\eqref{eq:qz}, of the selected map drops below a preselected threshold~\citep{heins2025axiomlearningplaygames}, which means that none of the currently in-play grounding likelihoods explain the data well. 
    
    \subsubsection{Clone-structured causal graphs}

    Clone-structured causal graphs (CSCG)~\citep{George2021} are a particular instance of Hidden Markov Models (HMM), where the observation likelihood maps observations deterministically to a multitude of states called "clones"; instead, state inference is driven entirely by the model's dynamics. The strength of clone graphs lies in the fact that, even though observations can be identical, this model can disambiguate them into distinct states. This approach has proven effective in navigation with aliased observations~\citep{George2021} and in hierarchical models of alternation tasks~\citep{VandeMaele2024}; moreover, the CSCG aligns closely with neural data on cognitive map formation in the hippocampus~\citep{sun2025learning}. Note that a clone graph with a single clone reduces to a standard (action-augmented) HMM.

    The CSCGs are learned using the expectation-maximization (EM) algorithm for HMMs (the Baum–Welch algorithm), which maximizes the evidence lower bound (ELBO)~\citep{George2021}. During the E-step, the posteriors over states are estimated through smoothing, i.e., the forward-backward algorithm. The M-step then selects the optimal parameters for the transition model, given this sequence of visited states. After training, the model is pruned using a Viterbi decoding. Here, for each timestep, the maximum likelihood state is selected, and the transition model parameters are estimated using these maximum likelihood states.

    We also implement a mechanism to learn the CSCGs online in Section~\ref{sec:ABCB}. In contrast to standard parameter learning in active inference, where the belief over the state is filtered as actions are executed and observations come in, clone-graphs smooth the belief over states, and propagate disambiguating information back into other ``cloned'' states. This provides a better estimate of the individual states, which can then be used to update the Dirichlet distribution over the transition parameters. For learning the schemas (i.e., the Level 2 transitions), we use a sliding window of 10 observations and update the parameters as each observation comes in. 
    
    We use the CSCG framework in three parts of this study. First, we use it to learn a cognitive map of the spatial locations in the environment. The clone-structure allows finding structure in highly ambiguous observations (6 tile colors, in 105 distinct locations).  Second, we use it in Simulation 1 to develop Level 2 of the HAI-$i$ agents without schemas that address the ABCD task. In this case, the CSCG is initialized with a number of clones $i$ that equals the number of blocks to be learned (20 or 40), to ensure the agent has enough capacity to learn all of them. Finally, we use the CSCG framework in Simulation 2 to develop Level 2 of the schema-based S-HAI-2C agent that addresses the ABCB task. In this case, the CSCG is initialized with 2 clones. To use them in discrete-time active inference, we need to map CSCGs to POMDPs as was described in~\cite{vandemaele2023integratingcognitivemaplearning}.
    



\section*{Code availability}

We provide the code for replicating the simulations at \url{https://github.com/toonvdm/grounding-schemas}.

\section*{Acknowledgments}

We thank Mohamady El-Gaby and Adam Harris for useful feedback and discussions. This research received funding from the European Research Council under the Grant Agreement No. 820213 (ThinkAhead), the Italian National Recovery and Resilience Plan (NRRP), M4C2, funded by the European Union – NextGenerationEU (Project IR0000011, CUP B51E22000150006, `EBRAINS-Italy'; Project PE0000013, `FAIR'; Project PE0000006, `MNESYS'), and the Ministry of University and Research, PRIN PNRR P20224FESY and PRIN 20229Z7M8N. The GEFORCE Quadro RTX6000 and Titan GPU cards used for this research were donated by the NVIDIA Corporation. We used a Generative AI model to correct typographical errors and edit language for clarity.

\newpage

\bibliography{references}
\bibliographystyle{unsrtnat}

\clearpage
\appendix

\setcounter{figure}{0} 
\renewcommand{\thefigure}{S\arabic{figure}}
\renewcommand{\theHfigure}{S\arabic{figure}}
\setcounter{table}{0} 
\renewcommand{\thetable}{S\arabic{table}}

\section{Supplementary Materials}

\subsection{Significance Tables for the ABCD and ABCB Tasks.}

These results ran 40 seeds for the sequential environment blocks. The independent statistical unit is the environment block: one randomly generated maze layout with its own reward sequence. Each agent is evaluated on the same 40 blocks (identical seeds across conditions), and all within-block measurements (i.e., cumulative reward over 10 000 steps, and steps-per-trial averaged over the trials in the block) are aggregated to a single value per block before testing, so $n=40$. Because the blocks are shared across conditions, comparisons use a paired t-test (\texttt{scipy.stats.ttest\_rel}) with the block as the pairing unit, which controls for per-maze difficulty.

\begin{table}[h!]
\centering 
\caption{\textbf{Effect sizes for Steps per Trial in the ABCD Task}. Paired comparisons over the n~=~40 environment blocks, which are the independent statistical unit: each agent is run on the same blocks, so the difference is taken within a block. $\Delta$ is the mean paired difference in the units of the read-out, with a 95\% confidence interval (\texttt{scipy.stats.ttest\_rel}); $d_z$ is Cohen's $d_z$ for paired samples. (*) marks $p < 0.05$. The corresponding box plot is depicted in Figure~\ref{fig:timesboxplot_abcd}.}
\begin{tabular}{llrrrl}
	\toprule 
	\textbf{Model A} & \textbf{Model B} & $\Delta$ (A$-$B) & \textbf{95\% CI} & $d_z$ & \textbf{$p$} \\
	\midrule 
	Random & HAI-20 & 168 & [142, 195] & 2.05 & $1.06 \cdot 10^{-15}$ (*) \\
	Random & HAI-40 & 236 & [218, 255] & 4.01 & $7.55 \cdot 10^{-26}$ (*) \\
	Random & S-HAI L & 249 & [233, 264] & 5.11 & $8.88 \cdot 10^{-30}$ (*) \\
	Random & S-HAI K & 249 & [233, 265] & 5.11 & $8.92 \cdot 10^{-30}$ (*) \\
	HAI-20 & HAI-40 & 68.1 & [41.7, 94.6] & 0.82 & $6.59 \cdot 10^{-6}$ (*) \\
	HAI-20 & S-HAI L & 80.5 & [55.5, 106] & 1.03 & $1.03 \cdot 10^{-7}$ (*) \\
	HAI-20 & S-HAI K & 80.7 & [55.7, 106] & 1.03 & $9.43 \cdot 10^{-8}$ (*) \\
	HAI-40 & S-HAI L & 12.4 & [-2.73, 27.5] & 0.26 & $1.05 \cdot 10^{-1}$ (n.s.) \\
	HAI-40 & S-HAI K & 12.6 & [-2.60, 27.8] & 0.26 & $1.02 \cdot 10^{-1}$ (n.s.) \\
	S-HAI L & S-HAI K & 0.17 & [-0.12, 0.47] & 0.19 & $2.40 \cdot 10^{-1}$ (n.s.) \\
	\bottomrule 
\end{tabular}
\label{tab:timesboxplot_abcd_effectsize}
\end{table}
\begin{table}[h!]
\centering 
\caption{\textbf{Effect sizes for Cumulative Reward in the ABCD Task}. Paired comparisons over the n~=~40 environment blocks, which are the independent statistical unit: each agent is run on the same blocks, so the difference is taken within a block. $\Delta$ is the mean paired difference in the units of the read-out, with a 95\% confidence interval (\texttt{scipy.stats.ttest\_rel}); $d_z$ is Cohen's $d_z$ for paired samples. (*) marks $p < 0.05$. The corresponding box plot is depicted in Figure~\ref{fig:rewardsboxplot_abcd}.}
\begin{tabular}{llrrrl}
	\toprule 
	\textbf{Model A} & \textbf{Model B} & $\Delta$ (A$-$B) & \textbf{95\% CI} & $d_z$ & \textbf{$p$} \\
	\midrule 
	Random & HAI-20 & -555 & [-742, -368] & -0.95 & $5.06 \cdot 10^{-7}$ (*) \\
	Random & HAI-40 & -1029 & [-1150, -907] & -2.71 & $9.81 \cdot 10^{-20}$ (*) \\
	Random & S-HAI L & -1056 & [-1138, -974] & -4.12 & $2.84 \cdot 10^{-26}$ (*) \\
	Random & S-HAI K & -1061 & [-1143, -979] & -4.14 & $2.27 \cdot 10^{-26}$ (*) \\
	HAI-20 & HAI-40 & -474 & [-655, -293] & -0.84 & $4.85 \cdot 10^{-6}$ (*) \\
	HAI-20 & S-HAI L & -501 & [-668, -333] & -0.96 & $4.48 \cdot 10^{-7}$ (*) \\
	HAI-20 & S-HAI K & -506 & [-672, -340] & -0.97 & $3.09 \cdot 10^{-7}$ (*) \\
	HAI-40 & S-HAI L & -26.9 & [-97.2, 43.4] & -0.12 & $4.44 \cdot 10^{-1}$ (n.s.) \\
	HAI-40 & S-HAI K & -32.0 & [-102, 38.4] & -0.15 & $3.64 \cdot 10^{-1}$ (n.s.) \\
	S-HAI L & S-HAI K & -5.10 & [-15.4, 5.18] & -0.16 & $3.22 \cdot 10^{-1}$ (n.s.) \\
	\bottomrule 
\end{tabular}
\label{tab:rewardsboxplot_abcd_effectsize}
\end{table}
\begin{table}[h!]
\centering 
\caption{\textbf{Effect sizes for Steps per Trial in the ABCB Task}. Paired comparisons over the n~=~40 environment blocks, which are the independent statistical unit: each agent is run on the same blocks, so the difference is taken within a block. $\Delta$ is the mean paired difference in the units of the read-out, with a 95\% confidence interval (\texttt{scipy.stats.ttest\_rel}); $d_z$ is Cohen's $d_z$ for paired samples. (*) marks $p < 0.05$. The corresponding box plot is depicted in Figure~\ref{fig:timesboxplot_abcb}.}
\begin{tabular}{llrrrl}
	\toprule 
	\textbf{Model A} & \textbf{Model B} & $\Delta$ (A$-$B) & \textbf{95\% CI} & $d_z$ & \textbf{$p$} \\
	\midrule 
	Random & S-HAI L & 241 & [220, 261] & 3.77 & $7.26 \cdot 10^{-25}$ (*) \\
	Random & S-HAI K & 242 & [222, 263] & 3.79 & $6.23 \cdot 10^{-25}$ (*) \\
	Random & S-HAI-2C L & 254 & [233, 275] & 3.82 & $4.37 \cdot 10^{-25}$ (*) \\
	Random & S-HAI-2C K & 255 & [234, 276] & 3.87 & $2.84 \cdot 10^{-25}$ (*) \\
	S-HAI L & S-HAI K & 1.31 & [-0.48, 3.10] & 0.23 & $1.47 \cdot 10^{-1}$ (n.s.) \\
	S-HAI L & S-HAI-2C L & 13.2 & [11.0, 15.3] & 1.96 & $4.14 \cdot 10^{-15}$ (*) \\
	S-HAI L & S-HAI-2C K & 14.3 & [12.1, 16.6] & 2.04 & $1.26 \cdot 10^{-15}$ (*) \\
	S-HAI K & S-HAI-2C L & 11.9 & [8.81, 14.9] & 1.24 & $1.44 \cdot 10^{-9}$ (*) \\
	S-HAI K & S-HAI-2C K & 13.0 & [11.0, 15.0] & 2.08 & $6.01 \cdot 10^{-16}$ (*) \\
	S-HAI-2C L & S-HAI-2C K & 1.17 & [-0.63, 2.97] & 0.21 & $1.98 \cdot 10^{-1}$ (n.s.) \\
	\bottomrule 
\end{tabular}
\label{tab:timesboxplot_abcb_effectsize}
\end{table}
\begin{table}[h!]
\centering 
\caption{\textbf{Effect sizes for Cumulative Reward in the ABCB Task}. Paired comparisons over the n~=~40 environment blocks, which are the independent statistical unit: each agent is run on the same blocks, so the difference is taken within a block. $\Delta$ is the mean paired difference in the units of the read-out, with a 95\% confidence interval (\texttt{scipy.stats.ttest\_rel}); $d_z$ is Cohen's $d_z$ for paired samples. (*) marks $p < 0.05$. The corresponding box plot is depicted in Figure~\ref{fig:rewardsboxplot_abcb}.}
\begin{tabular}{llrrrl}
	\toprule 
	\textbf{Model A} & \textbf{Model B} & $\Delta$ (A$-$B) & \textbf{95\% CI} & $d_z$ & \textbf{$p$} \\
	\midrule 
	Random & S-HAI L & -749 & [-825, -674] & -3.18 & $3.24 \cdot 10^{-22}$ (*) \\
	Random & S-HAI K & -761 & [-830, -692] & -3.52 & $8.62 \cdot 10^{-24}$ (*) \\
	Random & S-HAI-2C L & -1136 & [-1270, -1001] & -2.70 & $1.08 \cdot 10^{-19}$ (*) \\
	Random & S-HAI-2C K & -1155 & [-1287, -1023] & -2.81 & $2.81 \cdot 10^{-20}$ (*) \\
	S-HAI L & S-HAI K & -11.6 & [-30.2, 7.00] & -0.20 & $2.15 \cdot 10^{-1}$ (n.s.) \\
	S-HAI L & S-HAI-2C L & -386 & [-474, -298] & -1.40 & $7.06 \cdot 10^{-11}$ (*) \\
	S-HAI L & S-HAI-2C K & -406 & [-494, -318] & -1.47 & $1.81 \cdot 10^{-11}$ (*) \\
	S-HAI K & S-HAI-2C L & -375 & [-465, -284] & -1.32 & $2.99 \cdot 10^{-10}$ (*) \\
	S-HAI K & S-HAI-2C K & -394 & [-482, -306] & -1.43 & $3.91 \cdot 10^{-11}$ (*) \\
	S-HAI-2C L & S-HAI-2C K & -19.4 & [-39.2, 0.40] & -0.31 & $5.46 \cdot 10^{-2}$ (n.s.) \\
	\bottomrule 
\end{tabular}
\label{tab:rewardsboxplot_abcb_effectsize}
\end{table}

\newpage

\newpage
\subsection{Mixture of Grounding Likelihoods} 
\label{app:mogl}

\begin{figure}[h!]
    \centering
    \begin{minipage}{\linewidth}
        \centering
       \begin{minipage}{0.45\linewidth}
            \includegraphics[width=0.95\linewidth]{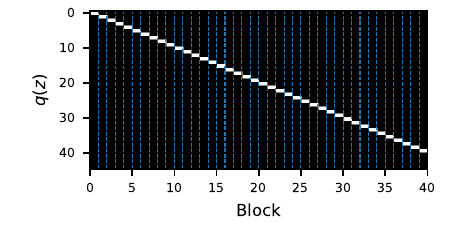} 
            \subcaption[]{}
       \end{minipage} 
       \begin{minipage}{0.45\linewidth}
            \includegraphics[width=0.95\linewidth]{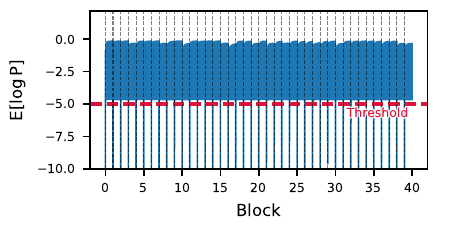} 
            \subcaption[]{}
       \end{minipage} 
    \end{minipage}
    
    \caption{\textbf{Mixture of Grounding Likelihoods.} (a) The approximate posterior over z (the
    selector variable of the grounding likelihood) during the ABCD task without repeated blocks, showing that the agent
    tends to select a different grounding likelihood for each of the 40 blocks. (b) The expected log likelihood during the
    40 blocks of the ABCD task without repeated blocks, with the threshold (-5) used for expanding the mixture model.
    \label{fig:mogl_norepeat} 
}
\end{figure}

We show the belief over the grounding likelihood and corresponding expected log likelihood curves for the sequential 40 blocks without repeating environment configurations in Figure~\ref{fig:mogl_norepeat}.

\subsection{Small Maze}
\label{app:smallmaze}

\begin{figure}[h!]
    \centering
    \begin{minipage}[b]{\textwidth}
        \centering
        \begin{minipage}[b]{0.48\textwidth}
        \centering
        \begin{minipage}{0.9\textwidth}
            \includegraphics[width=\textwidth]{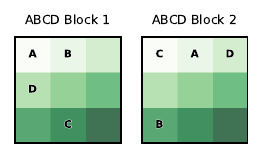}
            \subcaption[]{}
            \label{fig:small_abcd}
        \end{minipage}
        \begin{minipage}{\textwidth}
                \includegraphics[width=\textwidth]{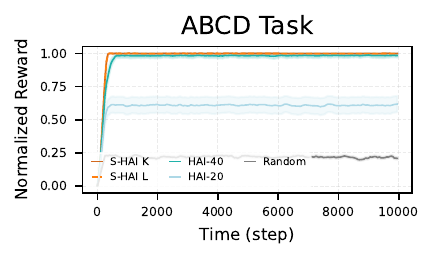} 
                \subcaption[]{}
        \end{minipage} 
        \begin{minipage}{\textwidth}
            \begin{minipage}{0.48\textwidth}
            \includegraphics[width=\textwidth]{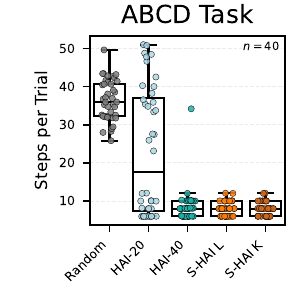}
            \subcaption[]{}
            \end{minipage} 
            \begin{minipage}{0.48\textwidth}
            \includegraphics[width=\textwidth]{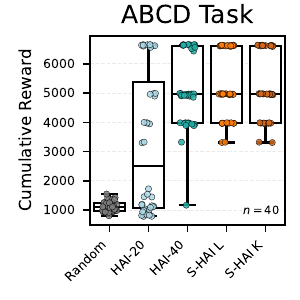}
            \subcaption[]{}
            \end{minipage} 
        \end{minipage}
        \end{minipage}
        \begin{minipage}[b]{0.48\textwidth}
            \centering
            \begin{minipage}{\textwidth}
                \centering
                \includegraphics[width=0.9\textwidth]{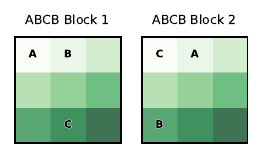}
                \subcaption[]{}
                \label{fig:small_abcb}
            \end{minipage} 
            \begin{minipage}{\textwidth}
                \includegraphics[width=\textwidth]{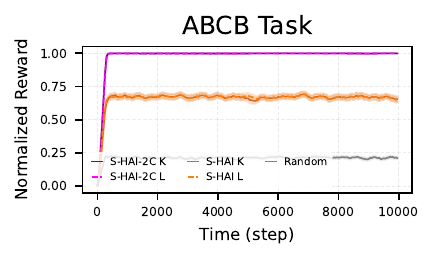} 
                \subcaption[]{}
            \end{minipage}
            \begin{minipage}{\textwidth}
            \begin{minipage}{0.48\textwidth}
                \includegraphics[width=\textwidth]{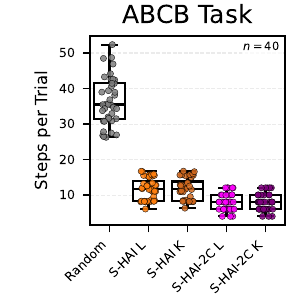}
                \subcaption[]{}
            \end{minipage}
            \begin{minipage}{0.48\textwidth}
                \includegraphics[width=\textwidth]{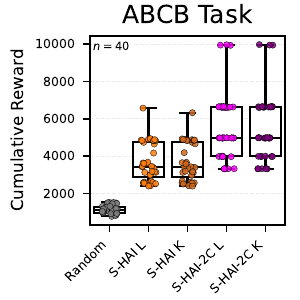}
                \subcaption[]{}
            \end{minipage}
            \end{minipage}
        \end{minipage}
    \end{minipage}
    \caption{\textbf{The small maze environment.} (a) The ABCD environment for two blocks. (b) Normalized reward in the ABCD task for the two schema-based agents using offline (S-HAI K) and online (S-HAI L) learning; two non-schema agents trained offline on 20 (HAI-20) or all 40 blocks (HAI-40); and the baseline (Random) agent. The solid line shows the mean normalized reward (measured as the reward rate smoothed over 250 steps, normalized against the optimal performance) across 40 blocks; the shaded area shows the standard error. Each block ends when the agent completes 10 000 navigation steps. 
    (c) Box plot of the average number of steps per trial required to complete an ABCD task and obtain 4 consecutive rewards, for the various agents 
    (d) Box plot of cumulative rewards across 40 ABCD blocks, for the various models
    (e) The ABCB environment for two blocks. 
    (f) Performance in the ABCB task for the offline and online schema-based agent with clone graphs (S-HAI-2C K, S-HAI-2C L), without clone graphs schema-based agents (S-HAI K, S-HAI L), and the Random agent.
    (g) Box plot of the average number of steps per trial required to complete an ABCB task and obtain 4 consecutive rewards, for the various agents 
    (h) Box plot of cumulative rewards across 40 ABCB blocks, for the various models
    \label{fig:smallmaze}
    }
\end{figure}

In this section, we provide the results for simulations on the ABCD and ABCB tasks, where the spatial level is fully observed, i.e. the tile color uniquely corresponds to the location of the agent. See Figure~\ref{fig:smallmaze} for the environment and behavioral results, which exhibit the same qualitative patterns reported in the main text.  

\subsection{Acquisition-to-transfer advantage conferred by using a learned schema}
\label{app:a2t}

\begin{figure}
    \centering
    \begin{minipage}{0.7\textwidth}
        \includegraphics[width=\textwidth]{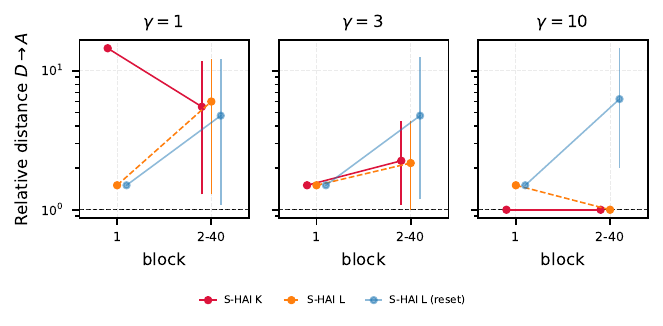}
        \subcaption[]{}
    \end{minipage}
    \caption{\textbf{Acquisition to transfer} (a) Relative distance (steps divided by shortest path; log scale, dashed line = optimal) for the first $D \rightarrow A$ transition per block, grouped for the first block and the last 39 blocks. S-HAI K has a pretrained schema, S-HAI L learns the schema online, S-HAI L (reset) does not retain the learned schema after each block. This experiment is executed for three variants of $\gamma$, where large value indicates more utility-driven behavior.}
    \label{fig:acquisition_to_transfer}
\end{figure}

In this experiment, we aim to investigate the advantage of using a schema when encountering a new configuration of a known task (i.e., an acquisition-to-transfer advantage). For this reason, we focus on the most informative transition: the D$\rightarrow$A transition after the first cycle of the task. This transition indicates whether the agent has learned the schema: although finding the A, B, C, and D goals initially requires exploration, once the agent has learned the ABCD schema, it can directly take the shortest path to A. 

Figure~\ref{fig:acquisition_to_transfer} shows the number of normalized steps (normalized by the shortest distance between two goals) for the fourth-to-fifth reward transition (i.e., $D \rightarrow A$) in each block. We consider three agent variations: (i) S-HAI K, an agent with a known schema whose only source of ambiguity is the mapping between the schema states and the physical locations. (ii) S-HAI L, an agent that has to learn both the schema and the mapping. Note that we do not reset the belief over state ($q(s^2_t)$) in between the blocks, and (iii) S-HAI L (reset) where the schema has to be relearned after every block. 

We consider three conditions for the experiment, based on the sampling precision for the grounding likelihood ($\gamma=\{1, 3, 10\}$). This value reflects the level of confidence the agent has in each observation. Lower values drive the agent more towards exploration, while high values yields a utility-driven agent. When learning both schema and grounding likelihood at the same time, a lower value for both $\gamma$ and the concentration parameter for the transition tensor are preferable, as this ensures that the agent acquires more evidence before fully committing to a grounding pair. 

The acquisition-to-transfer advantage of a learned schema is most prominently visible with a utility-driven agent, i.e. $\gamma=10$. We see that for the S-HAI K agent, the normalized distance is consistently 1 as the schema is already known. However, for the S-HAI L agent, in block 0, the agent still has a larger distance as it needs to discover the closing of the loop within the schema. In contrast, S-HAI L (reset) never reaches the optimum at $\gamma=10$, as it has to relearn the underlying schema at each block. Note that it takes the same time for S-HAI L (reset) as for S-HAI L for the first block, as finding the fifth reward requires the same level of exploration. At smaller values of $\gamma$, the exploratory behavior has the upper hand, and the benefit of the schema on time to acquire does not show as clearly. 

Note that block 1 is a single maze layout, whereas blocks 2-40 are averaged over 39 layouts. The block-1 value therefore reflects the geometry of that specific layout (e.g., how close goal A lies to the agent's exploration path), and at $\gamma=1$ and $\gamma=3$ it falls within the spread of the later blocks (Figure~\ref{fig:acquisition_to_transfer}). At these values of $\gamma$, we therefore observe no measurable transfer benefit on this metric; the benefit is visible only for the more utility-driven agent ($\gamma=10$).


\subsection{Quantification of schema generalization}
\label{app:seenunseen}

In order to better quantify the amount of generalization that schemas bring compared to learning the tasks in the models parameters, we separate the performance of seen (train) blocks with unseen blocks. 

\begin{figure}[h!]
    \centering
    \begin{minipage}{0.8\textwidth}
        \includegraphics[width=\textwidth]{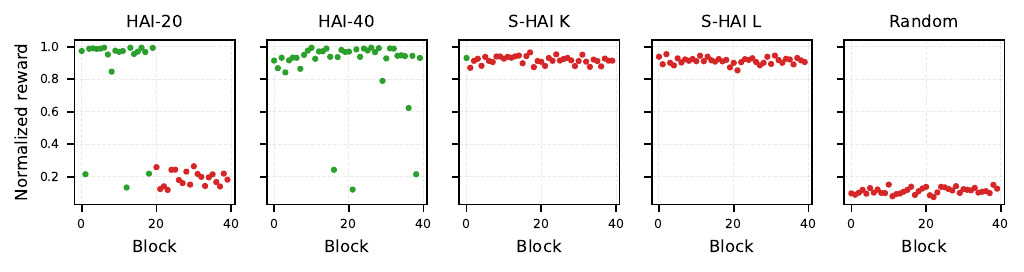}
        \subcaption[]{}
        \label{fig:cumulative_per_block}
    \end{minipage}
    \begin{minipage}{0.35\textwidth}
        \includegraphics[width=\textwidth]{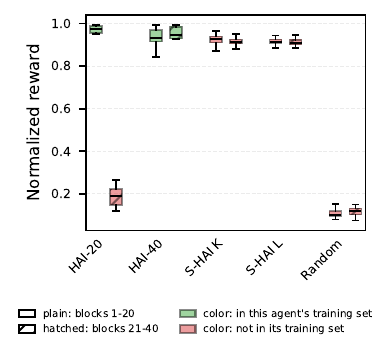}
        \subcaption[]{}
        \label{fig:cumulative_boxplot}
    \end{minipage}
    \begin{minipage}{0.35\textwidth}
        \includegraphics[width=\textwidth]{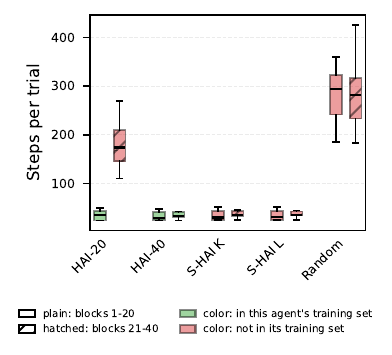}
        \subcaption[]{}
        \label{fig:step_boxplot}
    \end{minipage}
    \caption{\textbf{Generalization analysis on the ABCD task.} Comparing the HAI-20, HAI-40, S-HAI K, S-HAI L, and Random agents. (a) Normalized cumulative reward per block for the different agents. Green dots indicate that this configuration is part of the agent's training data, while red indicates an unseen block. (b) Box plot of the normalized cumulative reward for the various agents, pooled over the first 20 (plain), and last 20 (hatched) blocks. The color indicates whether this was part of the training set (green) or not (red). (c) Box plot of the steps per trial, pooled over the first 20 (plain), and last 20 (hatched) blocks. The color indicates whether this was part of the training set (green) or not (red).}
    \label{fig:seenunseen}
\end{figure}

Figure~\ref{fig:cumulative_per_block} displays the cumulative reward acquired per block. The green dots indicate blocks on which the agent is trained, while red dots indicate unseen scenarios. From this qualitative plot, it is clear that the HAI-i agents do not generalize beyond their training samples, while the S-HAI agents (both L and K) are able to reach similar highly rewarding performance, even though the blocks are unseen. 

Figures~\ref{fig:cumulative_boxplot} and~\ref{fig:step_boxplot} pool together these results and shows the cumulative reward and steps per trial respectively. Similar to before, performance for the HAI-20 and HAI-40 agents is highly correlated with the training data. HAI-20 performs well on almost all of the first 20 blocks, but shows performance below 0.4 on 3/20 blocks, while HAI-40 performs well on almost all 40 blocks, but shows performance below 0.4 and below 0.8 on 3/40 and 2/40 blocks, respectively. In contrast, S-HAI performs well across all blocks, even though these were not seen in training. 

\subsection{Statistics on the rate of convergence}
\label{app:convergence}

In order to quantify the speed at which the agent learns to solve the task, we measure the number of steps required to reach 95\% of the optimal reward rate for each block. We observe that when all structures in the block are novel, the mixture has no significant impact on performance, S-HAI MoGL L is indistinguishable from S-HAI L, and only the offline S-HAI MoGL K agent converges marginally slower. However, when block structures reoccur, both S-HAI MoGL agents converge much faster than their counterparts without a mixture. The statistics for the case without repeats are shown in Table~\ref{tab:stepstocriterion_abcd_effectsize}, the statistics for the case with repeats are shown in Table~\ref{tab:stepstocriterion_abcdrepeat_effectsize}.

\begin{table}[h!]
\centering 
\caption{\textbf{Effect sizes for Steps to 95\% of optimal in the ABCD}. Paired comparisons over the n~=~40 environment blocks, which are the independent statistical unit: each agent is run on the same blocks, so the difference is taken within a block. $\Delta$ is the mean paired difference in the units of the read-out, with a 95\% confidence interval (\texttt{scipy.stats.ttest\_rel}); $d_z$ is Cohen's $d_z$ for paired samples. (*) marks $p < 0.05$. }
\begin{tabular}{llrrrl}
	\toprule 
	\textbf{Model A} & \textbf{Model B} & $\Delta$ (A$-$B) & \textbf{95\% CI} & $d_z$ & \textbf{$p$} \\
	\midrule 
	S-HAI K & S-HAI L & -64.6 & [-146, 16.8] & -0.25 & $1.16 \cdot 10^{-1}$ (n.s.) \\
	S-HAI K & MoGL K & -140 & [-276, -3.04] & -0.33 & $4.53 \cdot 10^{-2}$ (*) \\
	S-HAI K & MoGL L & -65.7 & [-146, 14.9] & -0.26 & $1.07 \cdot 10^{-1}$ (n.s.) \\
	S-HAI L & MoGL K & -75.0 & [-199, 48.8] & -0.19 & $2.28 \cdot 10^{-1}$ (n.s.) \\
	S-HAI L & MoGL L & -1.02 & [-24.7, 22.6] & -0.01 & $9.31 \cdot 10^{-1}$ (n.s.) \\
	MoGL K & MoGL L & 73.9 & [-51.7, 200] & 0.19 & $2.41 \cdot 10^{-1}$ (n.s.) \\
	\bottomrule 
\end{tabular}
\label{tab:stepstocriterion_abcd_effectsize}
\end{table}
\begin{table}[h!]
\centering 
\caption{\textbf{Effect sizes for Steps to 95\% of optimal in the ABCD (repeat)}. Paired comparisons over the n~=~40 environment blocks, which are the independent statistical unit: each agent is run on the same blocks, so the difference is taken within a block. $\Delta$ is the mean paired difference in the units of the read-out, with a 95\% confidence interval (\texttt{scipy.stats.ttest\_rel}); $d_z$ is Cohen's $d_z$ for paired samples. (*) marks $p < 0.05$.}
\begin{tabular}{llrrrl}
	\toprule 
	\textbf{Model A} & \textbf{Model B} & $\Delta$ (A$-$B) & \textbf{95\% CI} & $d_z$ & \textbf{$p$} \\
	\midrule 
	S-HAI K & S-HAI L & -39.4 & [-93.7, 15.0] & -0.23 & $1.51 \cdot 10^{-1}$ (n.s.) \\
	S-HAI K & MoGL K & 498 & [295, 701] & 0.78 & $1.45 \cdot 10^{-5}$ (*) \\
	S-HAI K & MoGL L & 472 & [279, 665] & 0.78 & $1.49 \cdot 10^{-5}$ (*) \\
	S-HAI L & MoGL K & 537 & [324, 751] & 0.80 & $9.44 \cdot 10^{-6}$ (*) \\
	S-HAI L & MoGL L & 511 & [308, 715] & 0.80 & $9.80 \cdot 10^{-6}$ (*) \\
	MoGL K & MoGL L & -25.8 & [-107, 56.0] & -0.10 & $5.28 \cdot 10^{-1}$ (n.s.) \\
	\bottomrule 
\end{tabular}
\label{tab:stepstocriterion_abcdrepeat_effectsize}
\end{table}

\subsection{Qualitative Functional Correspondences}

Table~\ref{tab:funccorr} describes how each neural simulation in Figure~\ref{fig:neural_panel} is computed.

\begin{table}[h!]
\centering 
\caption{\textbf{Qualitative Functional Correspondences}}
\begin{tabular}{>{\raggedright\arraybackslash}p{0.2\linewidth} >{\raggedright\arraybackslash}p{0.3\linewidth} >{\raggedright\arraybackslash}p{0.4\linewidth}}
	\toprule 
	\textbf{Cell type} & \textbf{Represents} & \textbf{Computation} \\
	\midrule 
	Goal progress cells & Expected latent proximity to the current subgoal. & Expectation of the inductive matrix $I$ (see~\cite{friston2023active}) over the state belief at $t$, normalized to $[0,1]$ per block. \\
	\addlinespace
	Spatial firing field & The high-level state. & A goal state is the anchor, and the following goals in the schema sequence (found by rolling forward the transition model) are shown with fading intensity. \\
	\addlinespace
	Location-progress tuned & Position $\times$ expected proximity to the current subgoal. & Fires when the agent is at a specific location at a specific progress stage (early/mid/late). The angle shows which leg of the task the agent is on (e.g., A$\rightarrow$B) and how far along it is. \\
	\addlinespace
	Place cells & The low-level state. & Each visited location is colored by its most likely low-level state. \\
	\bottomrule 
\end{tabular}
\label{tab:funccorr}
\end{table}

\end{document}